\documentclass[manuscript]{aastex}
\def\lesssim{\mathrel{\hbox{\rlap{\hbox{\lower4pt\hbox{$\sim$}}}\hbox{$<$}}}}
\def\gtrsim{\mathrel{\hbox{\rlap{\hbox{\lower4pt\hbox{$\sim$}}}\hbox{$>$}}}}

\def\ltsima{$\;\buildrel < \over \sim \;$}
\def\simlt{\lower.5ex \hbox{\ltsima}}
\def\gtsima{$\;\buildrel > \over \sim \;$}
\def\simgt{\lower.5ex \hbox{\gtsima}}
\def\lesssim{\mathrel{\hbox{\rlap{\hbox{\lower4pt\hbox{$\sim$}}}\hbox{$<$}}}}
\def\gtrsim{\mathrel{\hbox{\rlap{\hbox{\lower4pt\hbox{$\sim$}}}\hbox{$>$}}}}
\def\gtrless{\mathrel{\hbox{\rlap{\hbox{\lower4pt\hbox{$<$}}}\hbox{$>$}}}}
\def\rightleftharpoons{\mathrel{\hbox{\rlap{\hbox{\raise2pt\hbox{$\rightharpoonup$}}}\hbox{$\leftharpoondown$}}}}
\def\notrightleftharpoons{\mathrel{\hbox{\rlap{\hbox{\raise1.5pt\hbox{$\;\mid$}}}\hbox{$\rightleftharpoons$}}}}
\def\dbar{\mathrel{\hbox{\rlap{\hbox{\raise3pt\hbox{$-$}}}\hbox{$d$}}}}
\def\hbar{\mathrel{\hbox{\rlap{\hbox{\raise3pt\hbox{$-$}}}\hbox{$h$}}}}
\def\nubar{\mathrel{\hbox{\rlap{\hbox{\raise2pt\hbox{$-$}}}\hbox{$\nu$}}}}
\def\lambdabar{\mathrel{\hbox{\rlap{\hbox{\raise2pt\hbox{$-$}}}\hbox{$\lambda$}}}}
\def\BbbV{\mathrel{\hbox{\rlap{\hbox{\raise2.5pt\hbox{${\rm v}$}}}\hbox{${\rm V}$}}}}
\def\BbbT{\mathrel{\hbox{\rlap{\hbox{\raise2pt\hbox{${\rm T}$}}}\hbox{${\rm T}$}}}}

\def\dddot{\hbox{\rlap{\hbox{\raise 8pt\hbox{${\bf ...}$}}}\hbox{$$}}}

\def\ltsima{$\;\buildrel < \over \sim \;$}
\def\simlt{\lower.5ex \hbox{\ltsima}}
\def\gtsima{$\;\buildrel > \over \sim \;$}
\def\simgt{\lower.5ex \hbox{\gtsima}}

\shorttitle{The Mass-Metallicity-Star-Formation Relation}
\shortauthors{Harwit and Brisbin}

\begin{abstract}
We describe an equilibrium model that links the metallicity of low-redshift galaxies to stellar evolution models.  It enables the testing of different stellar initial mass functions and metal yields against observed galaxy metallicities.  We show that the metallicities of more than 80,000 Sloan Digital Sky Survey (SDSS) galaxies in the low-redshift range $0.07\leq z\leq 0.3$ considerably constrain stellar evolution models that simultaneously relate galaxy stellar mass, metallicity, and star formation rates (SFRs) to the infall rate of low-metallicity extragalactic gas and outflow of enriched matter.  A feature of our model is that it encompasses both the active star forming phases of a galaxy and epochs during which the same galaxy may lie fallow.  We show that the galaxy-mass-metallicity-star-formation relation can be traced to infall of extragalactic gas mixing with native gas from host galaxies to form stars of observed metallicities, the most massive of which eject oxygen into extragalactic space.  Most consequential among our findings is that, on average, extragalactic infall accounts for one half of the gas required for star formation, a ratio that is remarkably constant across galaxies with stellar masses ranging at least from $M* = 2 \times 10^9$ to $6\times 10^{10} M_{\odot}$.  This leads us to propose that star formation is initiated when extragalactic infall roughly doubles the mass of marginally stable interstellar clouds.  The processes described may also account quantitatively for the metallicity of extragalactic space, though to check this the fraction of extragalactic baryons will need to be more firmly established.
\end{abstract}

\begin{document}

\title{Origin of the Galaxy Mass-Metallicity-Star-Formation Relation}
\author{Martin Harwit\altaffilmark{1,2} and Drew Brisbin\altaffilmark{3,1}}

\email{harwit@verizon.net}

\altaffiltext{1}{Center for Radiophysics and Space Research, Cornell University, Ithaca, NY, 14853}
\altaffiltext{2}{511 H street, SW, Washington, DC 20024-2725}
\altaffiltext{3}{National Radio Astronomy Observatory, Charlottesville, VA 22903, USA}

\section{Introduction}

The past decade has witnessed an explosion of observational studies of star formation enabled by powerful new observatories, active by now across the entire electromagnetic wavelength range from the gamma-ray region to the radio domain.  Theoretical studies of physical processes involved in star formation have also advanced. Despite these efforts, the physical conditions promoting star formation remain poorly understood.

Investigations currently pursued aim at understanding different aspects of the process.  Some researchers, among them \citet{Yate2012},  see contemporary star formation as part of a continuation of structure formation in the Universe. It begins with the condensation of massive haloes from a featureless plasma at earliest times. Baryonic matter flowing into these haloes clusters to form interacting galaxies, where stars are simultaneously forming as part of a grand gravitationally controlled process that continues to shape the Universe.  

Other investigators, e.g., \citet{Krum2012} pursue a more localized approach in search of the balance of thermodynamics, hydrodynamics, the interaction of matter with radiation, the forces of gravity, and chemical processes, all of which play a part in promoting or preventing the formation of stars in clouds of interstellar gases. 

The aim of the present paper is to search for new phenomenological clues to star formation and stellar evolution in the past few billion years that may be provided by Sloan Digital Sky Survey (SDSS) data on roughly $10^5$ well-characterized galaxies.  Phenomenological approaches have a rich history.  In a pair of seminal papers of 1998, \citet{Kenn1998a, Kenn1998b}, Rob Kennicutt pointed to two quite different features associated with, and potentially serving as measures of, the rate at which galaxies form stars.  The first, due to Maarten \citet{Schm1959}, was the column density of gas in galaxies exhibiting star formation.  The other, established by Kennicutt himself, was the H-$\alpha$ emission from known star-forming regions.  

That either of these features should provide a reasonable measure of the star formation rate (SFR) was remarkable. Both measures tacitly implied that star formation is a steady ongoing process, whereas a cursory glance at data gives the impression that star formation is episodic, punctuated by outbursts. Two circumstances, however, provide for a measure of stability.  The first is a shared initial mass function (IMF) originally defined by \citet{Salp1959}, who postulated that gravitationally bound associations of stars contained a well-defined distribution of stars of different masses. This IMF has  more recently been refined by \citet{Krou2001} and by \citet{Chab2003,Chab2005}, but remains central to all current thinking.  The second circumstance is that, despite its episodic appearance, star formation often continues for extended periods.  As new stars form, the more massive die out more rapidly but are replaced by younger equally massive stars as a region in which star formation began gradually shifts to neighboring domains.  

As \citet{Kenn1998b} explained, the observed H-$\alpha$ emission could be interpreted in two distinct ways. The first provided a measure of an instantaneous SFR:  Kennicutt noted that evolutionary synthesis models suggested that ``Only stars with masses of $>10 M_{\odot}$ and lifetimes of $<20$ Myr contribute significantly to the integrated ionizing flux, so the emission lines provide a nearly instantaneous measure of SFR, independent of the previous star formation history."  On the other hand, he also noted that, ``For integrated measurements of galaxies, it is usually appropriate to assume that the SFR has remained constant over time scales that are long compared with the lifetimes of the dominant UV emitting population ($<10^8$ year), in the `continuous star formation' approximation."  

These two distinct interpretations, as we shall see, do matter in discussions of the evidence provided by the SDSS in its census of low-redshift galaxies.  The past decade has shown that matter from the intergalactic medium continues to flow into galaxies at significant rates.  The influx may be episodic, at times rapidly initiating the formation of new stars,  at other times ceasing to do so.  But over the eons these sequences lasting up to hundreds of millions of years may be viewed as an ongoing history punctuated from time to time by some changes in course.   Star formation also leads to outflow, as the most massive stars formed explode as supernovae. These stars' ejecta are rich in metals and escape the galaxy to progressively raise the metallicity of the extragalactic medium.

As we will show in this paper, the SDSS reveals that, in the low-redshift universe, $z\leq 0.15$, low-mass galaxies with stellar masses $M_*$ in the range $1.8\times 10^9M_{\odot}\leq M_*\leq 7.1\times 10^{9} M_{\odot}$ exhibit  star formation in $\gtrsim 90\%$ of the galaxies observed, indicating that star formation is all but ubiquitous.  In contrast, for somewhat more massive galaxies with $M_* =  10^{10} M_{\odot}$ star formation is observed in $< 80\%$ of the galaxies, and the fraction drops to less than one half around $2 \times 10^{10} M_{\odot}$ with a further rapid decline at even higher masses where the appearance of galaxies switches from blue to red, and star formation declines or is quenched. 

\begin{figure}[t]
\centering
\includegraphics[width=13cm,trim=.5cm 1cm 5.cm 11cm, clip=true]{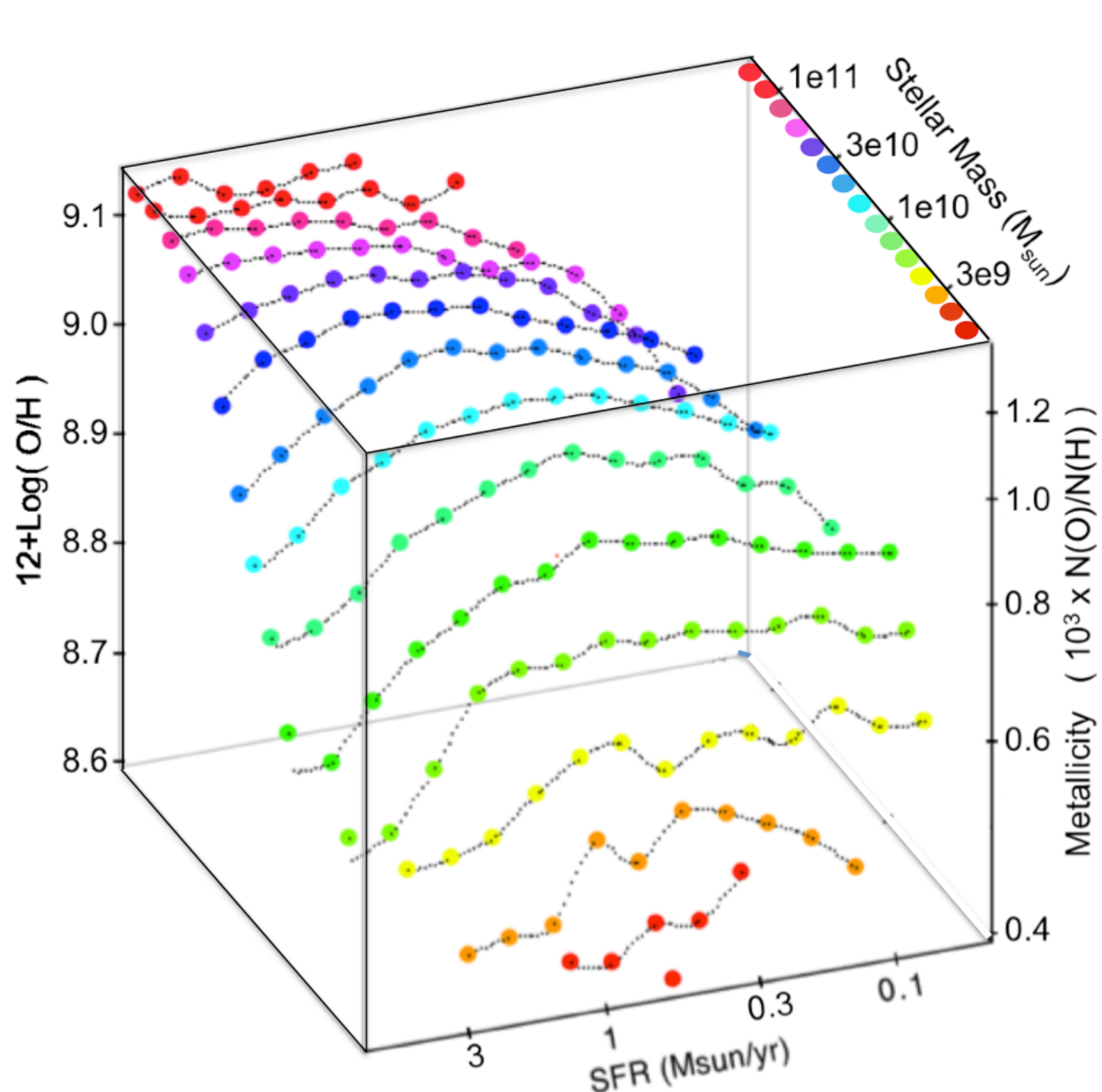}		
	\caption{Median metallicities for our sample of 81,583 SDSS galaxies, in the redshift range $0.07< z < 0.3$ as a function of the galaxies' stellar masses $M_*$ and SFRs in logarithmically designated $M_*$ and SFR bins comprising at least 50 galaxies in the indicated mass and SFR ranges.  The vertical axis establishes metallicity; horizontal axes respectively denote SFRs and $M_*$.  The sample shown includes galaxies from all our redshift and radius cuts described in Section 2.   Colored dots indicate SFRs derived from the data.  Dotted lines trace the relation between SFR and metallicity due to dilution by infalling matter, as discussed in Paper I.  The present figure, though founded on a somewhat different selection of galaxies, is based on a similar plot first published by \citet{Mann2010} and \citet{Lara2010}.  All our data and subsets of them divided by galaxy redshift and galaxy radius are available online.}
\label{fig:metallicities}
\end{figure}

Despite the existence of these two distinct galaxy populations --- a blue, low-mass population actively forming stars, and a red high-mass population all but devoid of star formation --- the same measures of SFR have usually been employed in discussions of both populations.  In Figure \ref{fig:metallicities}, one such SFR measure has been applied across the entire range of observed galaxies.   Whether this is appropriate is one of the questions raised in this paper.

A reason for concern is that a comparison by \citet{Torr2012} and more recently by \citet{WuYZ2013} shows that the ratio of nitrogen to oxygen abundances, [N/O], can be an order of magnitude higher in massive red galaxies than in blue low-mass galaxies.  The cause of this difference appears to be an accumulation of nitrogen preferentially expelled by intermediate mass stars roughly 0.4 Gyr after the high mass stars produced as part of the IMF have long ceased to exist.  An appreciable fraction of the H-$\alpha$ emission observed in these massive galaxies, as well as the observed [O III] emission, may thus be due to older intermediate-mass B0 to B8 stars rather than stars still actively forming.  If so, these traditional indicators of SFRs may bear little relation to current star forming activity at the time a massive galaxy is actually observed.  Rather it may indicate the existence of an intermediate mass stellar remnant of a star burst that had occurred $\sim 0.4$ Gyr earlier, but now no longer exists.  An early paper of \citet{Oppe2008}, hereafter O\&D(2008), explicitly takes this delayed metal enrichment of the interstellar medium into account.

Admittedly, then, no single measure of star formation rates might hold for a randomly selected galaxy viewed at an arbitrarily selected epoch.  But we may expect provisorily that when substantial numbers of galaxies are grouped by mass and size, as in Figure \ref{fig:metallicities}, or when any given galaxy is observed over its entire life, some set of shared characteristics will generally hold.  It has thus become customary to speak of a {\it mean shared history} of such galaxies that leads to roughly predictable measures of H-$\alpha$ emission indicative of the rate of star formation and to gas metallicities dependent on infall rates from extragalactic space.  That supernova explosions ejecting matter from galaxies must also be included in a complete picture has further been emphasized in a series of papers by Oppenheimer, Dav\'{e} and their co-authors, following the seminal paper of O\&D(2008). 
\clearpage

The large number of galaxies surveyed by the SDSS permits us to test the observational evidence for such a shared history, and may explain at least some aspects of the survey's findings.  In particular, there is considerable urgency to explaining the data for the redshift range $007\leq z< 0.3$ presented in Figure \ref{fig:metallicities}, in which galaxies appear embedded in a thin, curved, three-dimensional shell in a space defined by galaxy stellar mass $M_*$, oxygen metallicity $Z$, and SFR.  How can these galaxies adhere to this surface so closely when   \citet{Lovi2011} and \citet{Petr2012} provide clear evidence of high intragalactic metallicities in large clusters of galaxies; ram pressures that may strip a galaxy of its interstellar gases; potential increases in the metallicity of interstellar gases through capture of intracluster matter; ejection of the galaxies' gaseous contents through violent explosion; and many other factors?  

Despite such tumultuous intrusions the mean characteristics of the SDSS sample of galaxies remains remarkably well defined.  

The purpose of the present paper is to come to grips with how and why these apparent contradictions persist, and to make sense of them.  The model we develop may appear overly simplistic in view of the many obvious complexities just listed; but some simple design appears to be required to account for the clean geometric sheet displayed in Figure \ref{fig:metallicities}, and this is what we hope to uncover.  

In \citet{Bris2012}, hereafter referred to as Paper I, we were primarily concerned with finding an analytic relationship between mass infall rates and star formation rates, showing that infalling and native gas effectively mix before star formation sets in, and demonstrating that the infalling gas appears pristine, with essentially no sign of metal content.  A limitation of the paper, however, was the assumption that mass outflow through supernova ejection would have minor effects on metallicity and that the primary source of metals in galaxies would be contributed through delayed return from evolved stars.  In the current paper we take ejection of mass from galaxies into full account, and also consider the effects of fallow periods in which star formation is interrupted, a feature we believe has not yet been discussed by other authors.

The emphasis in the present paper is on a set of guiding principles which, if not rigorously obeyed, at least appear to hold well enough at low redshifts to provide an overview of the ways in which mass inflow, star formation, the production and ejection of metals by massive stars, the metallicity of interstellar gases, and outflow from galaxies combine to produce a self-consistent mean shared history.  Constraints explaining such a history, at least over periods of order 1 Gyr at low redshifts, are a combination of assumptions on, and partially verifying observations of: (i) a steady IMF independent of $M_*$, (ii) predictable stellar evolution and metal production rates, (ii) nearly constant gas metallicity, (iii) negligibly increasing galaxy masses over time, and (iv) averaged over long time intervals and large groups of galaxies having identical masses and identical metallicities, a nearly constant ratio of gas to stellar mass.  We examine each of these properties in some depth to show that the galaxy mass-metallicity-star-formation relation of Figure \ref{fig:metallicities} can be quantitatively understood in those terms, and that a number of predictions emerge, which may be observationally tested.

In Section 2 of our paper we provide the selection criteria we applied to cull out a sample of SDSS galaxies whose physical characteristics are reliably defined.    Section 3 lists a further set of criteria applied to more clearly establish the episodic nature or continuity of star forming epochs and to define four SDSS galaxy subsets by indicator of star-forming activity. Section 4 describes the distribution of SDSS galaxies by stellar mass $M_*$, star-formation rates SFR, and population density, i.e., numbers of galaxies found in different $M_*$ and SFR ranges.  In Sections 5  and 6 we introduce the metallicity distributions across these same $M_*$ and SFR ranges and then search for signs, respectively, of metallicity and mass evolution among low-redshift galaxies, on billion-year time scales, but find both to be too low to be significant in our discussions of this set of galaxies.  Section 7 considers the episodic nature of star formation and intervening fallow intervals on the distribution of galaxy population densities among our four star-forming subsets.  With these criteria established Section 8 outlines a toy model of equilibrium star formation.  In Section 9 we turn to the metallicity enrichment of the interstellar medium through stellar evolution and the influence of an initial mass function.  In Section 10 we test a specific stellar and galaxy evolution model described by O\&D(2008) against our toy model to determine whether it leads to oxygen metallicities consistent with those found in SDSS galaxies. We find that it roughly does. In Section 11 we note that the uniform metallicities at different H-$\alpha$ luminosities in high mass  galaxies provide ways to estimate mass loss to the extragalactic medium from lower-mass galaxies; we further show that infalling gas, on average, contributes half the gas leading to star formation.  A final Section 12 points to additional studies that can benefit from investigations along the lines pursued in the present paper, to provide answers to interesting questions based entirely on metallicity data and a reliance on continuity and conservation arguments, as contrasted to more difficult dynamical approaches that would require data currently not in hand. 

\section{Systematic Derivation of SDSS Galaxy Properties}

Our selection of galaxies comes from the MPA-JHU catalog compiled from SDSS data release 7 (DR7) available online at (http://www.mpa-garching.mpg.de/SDSS/DR7/).  

To permit closer comparison of our data and those of \citet{Mann2010,Yate2012} and Paper I, we used the DR7 rather than a later SDSS data release, and employed similar selection criteria as those in the three earlier studies.  These involved selections based on redshift, ($0.07\leq z< 0.3$), H-$\alpha$ signal to noise ratio ($\geq$10), and exclusion of galaxies exhibiting active galactic nuclei (AGN) using the criteria put forward by \citet{Kauf2003a}. Our stellar mass estimates were taken directly from the MPA-JHU catalog of \citet{Kauf2003b}, with a correction to convert from masses based on a Kroupa IMF, \citep{Krou2001}, to a Chabrier IMF, \citep{Chab2003}.  Star formation rates were determined using H-$\alpha$, based on the work of \citet{Kenn1998b} with a correction for a Chabrier IMF.  Although G. Chabrier later updated his initial mass function \citep{Chab2005} and found it in better agreement with current data, as indicated by \citet{Parr2011}, we chose to work with his earlier IMF in order to compare our observational data and analytic model with those of Paper I.  We restricted our final sample to sources with stellar masses between 1.06$\times$10$^9$ and 2.66$\times$10$^{11}$ M$_\odot$ and SFRs between 0.03 and 7.5 M$_\odot$/yr.

Because H-$\alpha$ measurements are made uncertain by foreground dust extinction, we corrected for this using the Balmer decrement, along the lines established by \citet{Card1989}.  We excluded any sources with extinction corrections A$_V > 2.5$.  

We used metallicities from the MPA-JHU catalog based on the \citet{Trem2004} method.  This method is based on Bayesian analysis of emission lines modeled within a grid of metallicities. A metallicity probability distribution is created for each source by comparing the intensities of observed emission lines with the grid of modeled emission lines. We adopt the mean of the probability distribution as the characteristic metallicity for our sources. This is in contrast with other common metallicity diagnostics which use a set of strong emission line ratios to determine metallicity. We refer interested readers to the online data accompanying Paper I, in which we used the R23 oxygen line diagnostic and the [NII] 6584/H-$\alpha$ ratio, both common metal indicators, to analyze the metallicity of our sample. In Section 12 we briefly discuss the robustness of our results with respect to the method used for determining metallicity. 
Most of the sources we studied fall into a galactic stellar mass range from $M_*\sim 1.77\times10^9$ to $5.64\times 10^{10} M_{\odot}$, with star-formation rates ranging roughly from $0.071$ to $6.4 M_{\odot}$ yr$^{-1}$, and metallicities determined by number of oxygen atoms relative to hydrogen, spanning a range from $Z\sim4\times 10^{-4}$ to $1.4\times 10^{-3}$, straddling a Solar System abundance of $Z\sim 8.5 \times 10^{-4}$. 

The theoretical treatment of our findings involves additive properties of metallicity, rather than metallicity ratios.  Accordingly, we express metallicities, throughout, in terms of actual oxygen abundances, as opposed to their logarithmic values more conventionally adopted. 

The redshift range covered both by \citet{Mann2010} and by us is set at a minimum value of $z = 0.07$ to ensure that the [OII]$\lambda3727$ emission line falls well within the useful spectral range, and that the 3 arc second aperture of the SDSS spectroscopic fiber will sample a significant fraction of a galaxy's surface area.  At $z = 0.07$, an aperture of 3 arc seconds, corresponds to a spatial diameter of 4 kpc implying that we probe only the central 2 kpc regions of a galaxy.  At $z = 0.3$, the aperture corresponds to a diameter of 13.2 kpc, and thus samples a larger fraction of the galaxy.  We make use of a Hubble constant $H_0 = 71$ km s$^{-1}$ Mpc$^{-1}$, $\Omega_{DE} = 0.73$ and $\Omega_M = 0.27$ throughout \citep{Wrig2006}.

In addition to these selection criteria imposed by \cite{Mann2010}, Paper I, and the present paper, we required our sample to have an observed Petrosian half-light radius in the {{$r$-color}} band, r$_{50}$.  This assured an ability, in the spirit of \citet{Elli2008}, to investigate the role that galactic radius might play in star formation and metallicity.  We also took pains to eliminate duplicate observations from our SDSS sample.  Where an object was observed multiple times, we averaged its properties across the multiple entries.

Both \citet{Mann2010} and \citet{Elli2008} treated galaxies in the redshift range we cover as though they were coeval.  To examine whether or not further insight could be gained through study of the provenance of the sample galaxies, we followed the thrust of Paper I by dividing the SDSS galaxies into three separate redshift ranges, $0.07\leq z < 0.10,\ 0.10\leq z< 0.15,\ {\rm and}\  0.15\leq z< 0.30$  --- hereafter, respectively, referred to as the low, medium, and high-redshift ranges --- and three Petrosian half-light radii r$_{50}$ ranges, small r$_{50} < 3.74$ kpc, medium 3.74$\leq$r$_{50} < 5.01$, or large r$_{50} \geq $5.01. Each of our tables reflecting this breakdown by redshift and radius divides galaxies according to logarithmic mass and logarithmic SFR. The full tables are available online and represent extensions of tables originally placed online as part of Paper I.

Although many of the entries in the tables we have archived online show galaxy populations that may be quite sparse in certain galaxy mass/SFR bins, we considered our findings significant only if based on  bins that have $\geq 50$ galaxies per bin.  The abridged tables presented in this paper conform to this criterion and are organized by logarithmic galaxy stellar mass ranges $M_*$ and logarithmic star formation rate ranges SFR($x$) using bins populated by $N(x) \geq 50$ galaxies, where the variable $x$ specifies star formation rates measured in solar masses per year, $M_{\odot}$ yr$^{-1}$.  As an alternative we could have considered requiring valid bins to satisfy a maximum dispersion in their average metallicity. This would have similarly emphasized bins with a large population of galaxies, but in order to compare directly to the work of \citet{Mann2010} and \citet{Yate2012} we used their identical 50 galaxies per bin minimum criterion. Any readers interested in independently investigating dispersion-based criteria should consult the online tables where the standard deviation of the metallicity is presented in tabular format for each bin.

While the SDSS provides a way of studying star formation statistically in a sample exceeding a hundred thousand galaxies, both our selection and those of  \citet{Mann2010}, \citet{Elli2008}, and Paper I restrict themselves to observations of the central portions of galaxies, in many of which spiral arms no longer are well defined.  Our studies thus are not adequate for addressing questions of spiral structure or its effects on star formation.  

Studies of H$_{\rm II}$ regions outside these central portions of nearby galaxies do have the potential of revealing such effects in greater detail.  \citet{Sanc2013} have undertaken such studies and point out a potential disagreement of their results with those of \citet{Mann2010}, \citet{Lara2010}, \citet{Yate2012}, and Paper I.  These four studies have highlighted a correlation of star formation rates with  metallicity, which \citet{Sanc2013} do not find.  This difference, however, is likely to be due to the preponderance of H$_{\rm II}$ regions in massive galaxies on which the \citet{Sanc2013} study is based. For massive galaxies this metallicity dependence is absent also in the SDSS galaxies; only the  lower mass SDSS galaxies clearly exhibit this dependence, and for these galaxies the sample studied by \citet{Sanc2013} is relatively small and exhibits scatter.  These differences, while not yet resolved to everyone's satisfaction do not appear to us to be pressing.  

Thus, even though our observations largely deal with the central portions of galaxies, this emphasis has an interest of its own given the information they may provide on these generally active regions.

\section{Selection of Star-Forming Galaxies from the SDSS}

\begin{figure}[h]
\centering
\includegraphics[width=16cm,trim=0cm 0cm 0.cm 0cm, clip=true]{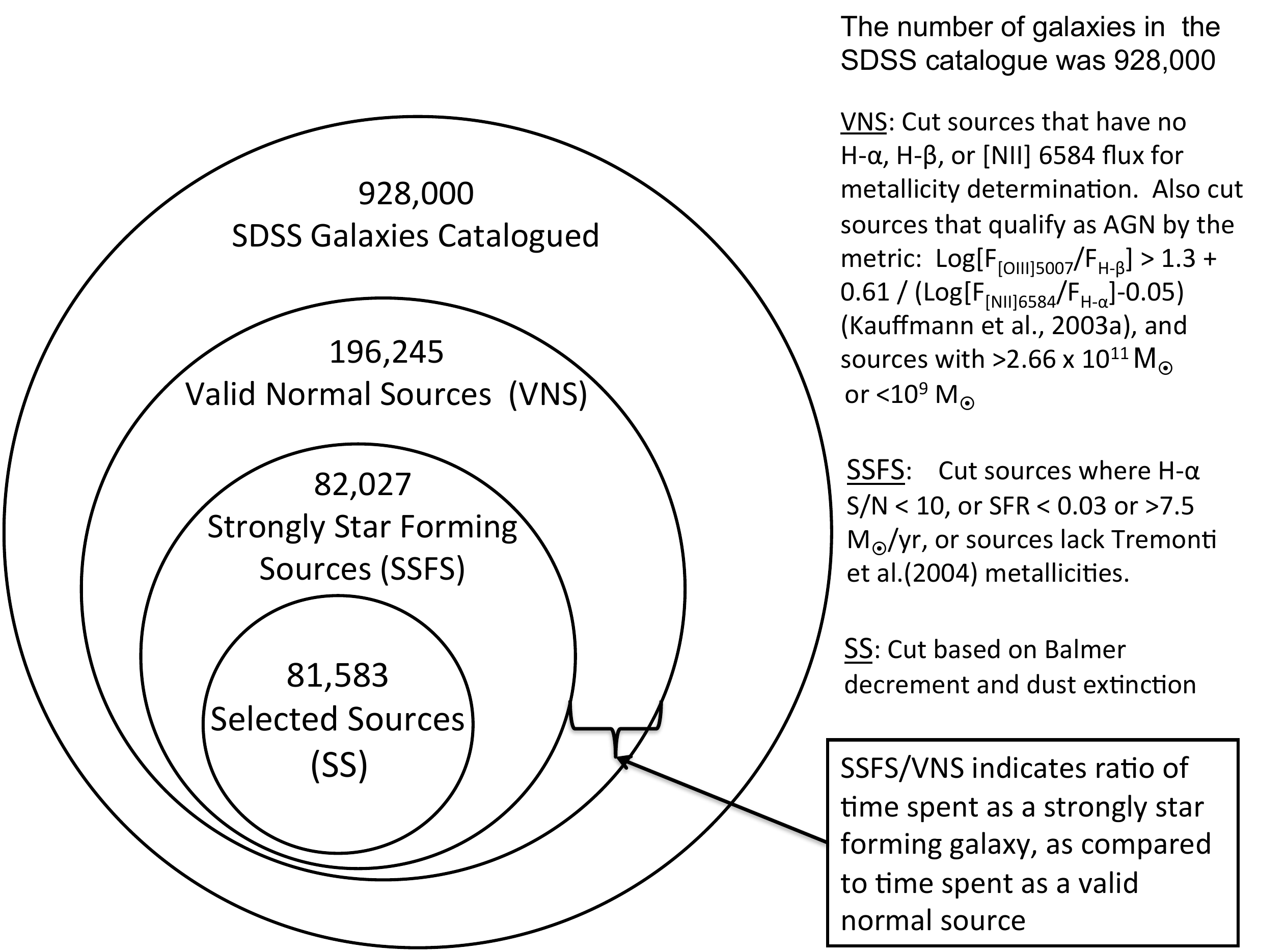}
\caption{Selection criteria applied to various samples of galaxies extracted from the SDSS Survey.  The criteria are further explained in the text.}
\label{fig:selection}
\end{figure}

A major purpose of the present paper is to demonstrate that star-forming galaxies undergo distinct evolutionary phases, which cannot be neglected if a coherent picture of star formation is to emerge.  These phases are characterized by different observable traits, which we first define.  Figure \ref{fig:selection} provides a breakdown of traits exhibited by $\sim 928,000$ SDSS galaxies.  

i) From this list of galaxies we restricted ourselves to the limited redshift range $0.07 < z < 0.3$ containing a sufficiently large galaxy population with accessible emission lines to potentially support significant conclusions. To gain insight on a relatively uniform set of star forming galaxies, we further eliminated any sources whose line spectra indicated the presence of a dominant active galactic nucleus (AGN). For this we employed the criteria established by \citet{Kauf2003a}, identifying purely star forming galaxies as having Log([OIII]5007/H-$\beta$)$<$1.3+0.61/(Log([NII]6584/H-$\alpha$)-0.05) and Log([NII]6584/H-$\alpha$)$<$0.05. To obtain a measure of metallicity as well as dust extinction for this sample we eliminated galaxies in which the emission lines H-$\alpha$, H-$\beta$, and [NII] 6584, key to determining metallicity using the \citet{Trem2004} method, were undetected. We further restricted our sample to include sources within the stellar mass range discussed in Section 2. This pared our sample down to $\sim$196,245 galaxies we label in Figure \ref{fig:selection} as \textit{Valid Normal Sources} (VNS).

\clearpage

ii) To further ensure the quality of our data, particularly our derived SFRs, we culled out sources with H-$\alpha$ signal-to-noise ratios below 10. We also culled any sources which had undetermined metallicities based on the \citet{Trem2004} method, as well as those with SFRs beyond our adopted range of 0.03 to 7.50 M$\odot$ year$^{-1}$ capturing the majority of SDSS galaxies.
This left us with 82,027 galaxies we call \textit{Strongly Star Forming Sources} (SSFS).
  
iii) The SSFS sample represents sources that can reliably be taken to be currently forming stars. To ensure robust metallicity estimates, however, we made two additional cuts. Any sources with a Balmer decrement below 2.5 or visual extinction 
magnitude in excess of 2.5 were eliminated.
This final cut only eliminated 0.5\% of the SSFS galaxies, suggesting the properties of the larger SSFS sample are already quite robust. This yields a set of 81,583 galaxies constituting our most stringently defined \textit{Selected Sources} (SS).


\section{Relations Between Star Formation, Metallicity and Population Density}
\label{section:Metallicity}
\begin{figure}[h!]
	\centering
\includegraphics[height=1\textwidth,angle=90,trim=3.5cm 1cm .75cm 1.5cm, clip=true]{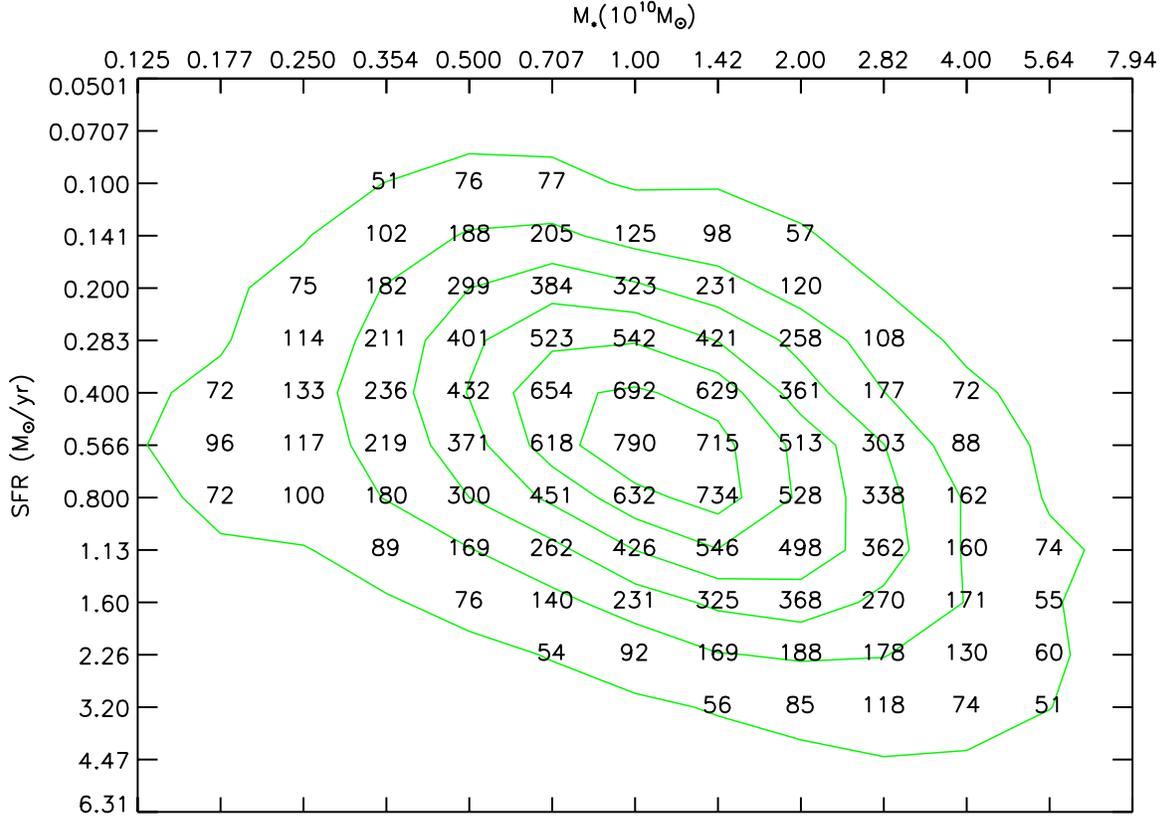}	
\caption{Low-Redshift Small-Radius (SSFS) Populations in Bins Comprising $\geq 50$ Galaxies. Contours represent intervals of 125 with the bottom contour at 50.}
\label{table:SSFS-Populations}
\end{figure}

We now turn to some of the key relations among SS, SSFS and VNS galaxies regarding the sample of low redshift, small radius galaxies, which is the most richly sampled subset of our SDSS sample. Our online tables characterize the sample sizes, mean metallicities, and metallicity standard deviations, for all redshift and radial size subsets of our data, broken down by star formation rate and stellar mass. The first set of low-z-small-r data, shown in Figure \ref{table:SSFS-Populations}, exhibit the population density distribution of SSFS galaxies in terms of galaxy stellar mass $M_*$ and star formation rate SFR. The peak in the distribution is found among galaxies of $M_* = 1.0\ {\rm to\ 1.4}\times 10^{10} M_{\odot}$ at star formation rates, ${\rm SFR} \sim 0.5$ to 0.80 $M_{\odot}$ yr$^{-1}$.


The population distribution among Strongly Star Forming Galaxies, SSFS, in Figure \ref{table:SSFS-Populations} shows close similarity to the SS distribution. 
This is clearly indicated by the ratio of these populations shown in Table \ref{table:SFSS/VNS-Ratios}, which is close to unity everywhere, but begins to diverge for high-mass galaxies.



Moreover, as Table \ref{table:SFSS/VNS-Ratios} shows, SSFS galaxies form a major subset of the VNS grouping as seen in a comparison of population bins containing $\geq 50$ galaxies in the various $M_*$ ranges covered at the respective redshifts.  In Figure \ref{table:SSFS-Populations}, the peak population over the logarithmically spaced bins is fairly flat over star forming rates from 0.28 to 0.8 $M_{\odot}$ yr$^{-1}$ and only slightly narrower in the $M_*$ range covered. In both tables, a high-density ridge in the distribution dips slightly from upper left to lower right, hinting at the well known {\it galaxy main sequence} discussed by  \citet{Elba2011}, who studied the far infrared emission from these galaxies using Herschel and Spitzer data, and identified different contributors to their galaxy sets through differences in their respective ultraviolet and X-ray emissions.  In our data, the galaxy main sequence is readily apparent when our entire set of galaxies in the range $0.07\lesssim z < 0.3$ is displayed in unison.  But, as we pointed out in Paper I, dividing this range into its low, medium, and high-redshift components shows that low-mass galaxies populate the main sequence mainly at low star formation rates and low redshifts; intermediate and high-mass galaxies, respectively, appear at progressively higher SFRs and higher redshifts.  The full extent of the main sequence is thus not readily discerned in narrow redshift slices.


\begin{table}[t]
\caption{Small-radius, low- and medium-redshift (SS), (SSFS) and (VNS) Population Ratios}
\vskip0pt\vskip-24pt\vskip0pt
\begin{center}
\begin{tabular}{lccccccccccl} \hline\noalign{\vskip3pt}
$M_*(10^{10}M_{\odot})$&0.177&0.250&0.354&0.500&0.707&1.00&1.42&2.00&2.82&4.00&5.64\\
 \hline\noalign{\vskip1pt}
Low-z SS& 451  &     730  &   1365   & 2394  &     3427   &   3955  &    4013   &   3053 &    1991  &   967  &    367\\
Low-z SFSS & 451  &      730   &   1367  &    2394  &    3428  &   3957 &    4014   &   3070   &   2014   &   988  &     392\\
Low-z VNS    & 465 &  745  &    1383   &   2459   &   3788   &   5016   &   6045  &    5915   &   4761  &    2966    &   1697\\
$N_{{\rm SS}}/N_{{\rm SSFS}}$&1.00 &1.00 &  1.00&  1.00 &  1.00 &  1.00 &  1.00 &  0.99 &  0.99  & 0.98 &  0.94\\
$N_{{\rm SSFS}}/N_{{\rm VNS}}$& 0.97 & 0.98 &  0.99 &  0.97 &  0.90 &  0.79 &  0.66 &  0.52&  0.42&  0.33 &  0.23\\ 
 \hline\noalign{\vskip1pt} 
Med-z SS&&   309    &   510  &    723   &    834  &   995    &  1273   &   1560   &   1429   & 1012   & 421\\
Med-z SSFS&&    309   &    512   &    723  &    834    &  995   &   1274   &   1562  &    1436  & 1023   & 431\\
Med-zVNS&&    332    &   538   &    752   &  883   &   1076    &  1425    &  2001   &   2575  & 3086 & 2857  \\
$N_{{\rm SS}}/N_{{\rm SSFS}}$&&1.00 &  1.00 &  1.00 &  1.00 &  1.00 &  1.00 &  1.00&  1.00 &  0.99 &  0.98\\
$N_{{\rm SSFS}}/N_{{\rm VNS}}$&&0.93 &  0.95 &  0.96 &  0.94 &  0.92&  0.89 &  0.78 &  0.56 &  0.33 &  0.15\\
\hline\noalign{\vskip1pt}
\end{tabular}
\end{center}
\label{table:SFSS/VNS-Ratios}
\end{table}

\section{Galaxy Metallicity Evolution at Low Redshifts}\label{section:metallicity-evolution}

\begin{figure}[h!]
	\centering
\includegraphics[height=1\textwidth,angle=90,,trim=3.5cm 1cm .75cm 1.5cm, clip=true]{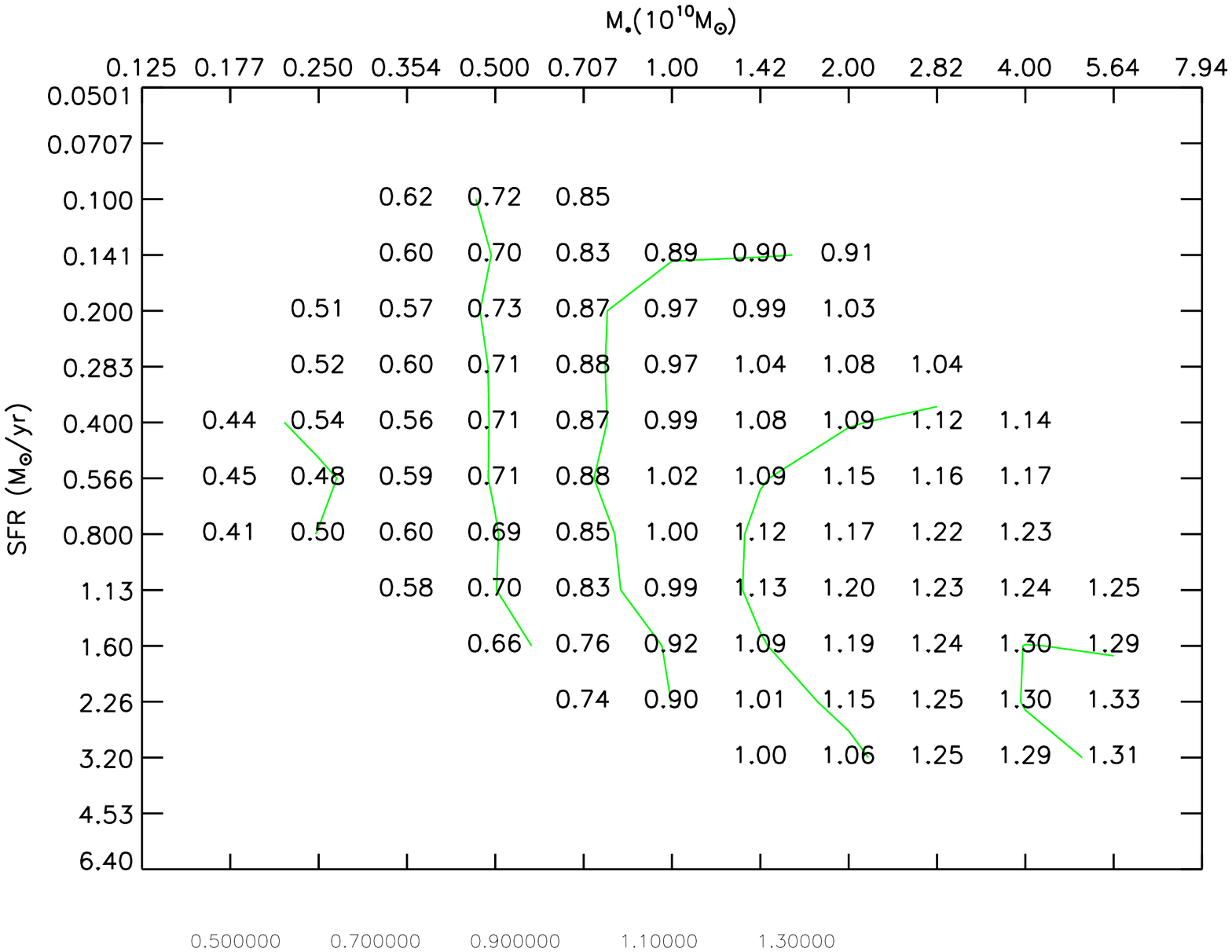}
\caption{Metallicities (in units of 10$^{-3}$ (O/H)) in Bins Comprising $\geq 50$ Low-z Small-Radius (SSFS) Galaxies. Contours represent intervals of 0.2 with the bottom contour at 0.5.}
\label{table:SSFS-metallicity-Low-z}
\end{figure}



\begin{figure}[h!]
	\centering
\includegraphics[height=1\textwidth,angle=90,trim=3.5cm 1cm .75cm 1.5cm, clip=true, clip=true]{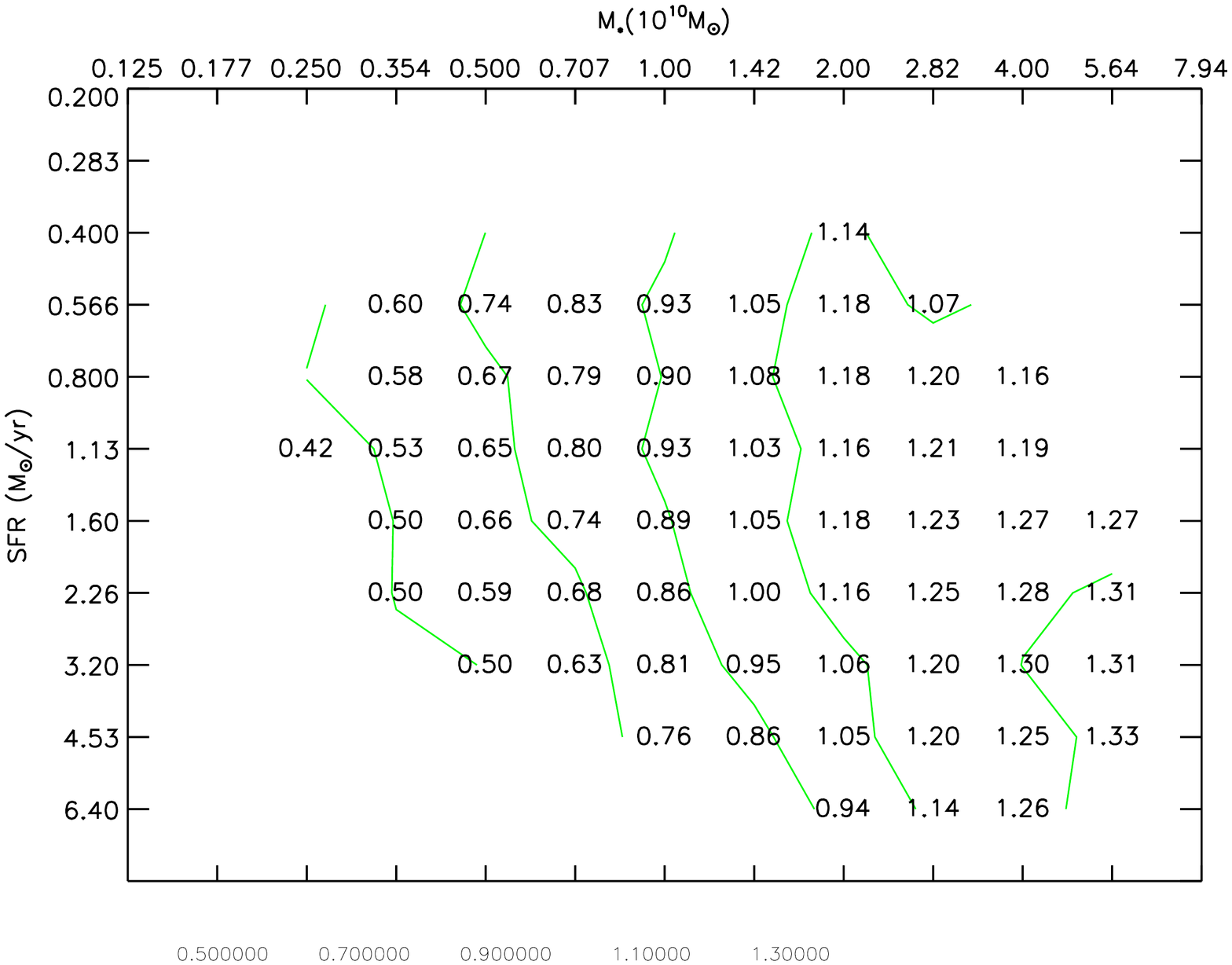}
\caption{Metallicities (in units of 10$^{-3}$ (O/H)) in Bins Comprising $\geq 50$ Medium-z Small-Radius SSFS Galaxies. Contours represent intervals of 0.2 with the bottom contour at 0.5.}
\label{table:SSFS-metallicity-Med-z}
\end{figure}

We find that, for random variations of the metallicity in galaxies of arbitrary mass $M_*$ producing stars at a given rate ${\rm SFR}(x)$, the standard deviation from the mean metallicity values cited in Figure \ref{table:SSFS-metallicity-Low-z} is of order 0.15 to $0.3\times 10^{-3}$. This means that, for our minimum bin populations of 50 galaxies per bin, the standard error of the mean metallicities should generally be no greater than $\sigma(Z_x)\sim 0.028\times 10^{-3}$.  In bins with populations $\geq 500$ in Figure \ref{table:SSFS-Populations}, the standard error of the mean should be well within $\sigma(Z_x)\sim 0.01\times 10^{-3}$.

An instructive comparison comes from metallicity data at low and medium redshifts, respectively in the ranges $0.07\leq z < 0.10$ and  $0.10\leq z<0.15$. Figure \ref{table:SSFS-metallicity-Med-z} provides these data for ${\rm (SFR})/M_*$ bins containing $\geq 50$ SS galaxies in the medium-redshift range.  As Table \ref{table:metallicity-evolution} shows, this still provides 50 bins, for a total of 6592 galaxies for which comparisons can be made. The SFR ranges in common to both tables understandably are restricted to higher SFRs, and more massive galaxies where data on the more distant galaxies are more reliably obtained.

Although this comparison deals with two sets of galaxies observed at different epochs, we know that we are at least dealing with galaxies of essentially identical mass, because, as we will show in Section \ref{section:rates}, below, at the observed rates of star formation the stellar masses of these galaxies cannot appreciably change over a period of $\sim 0.5$ Gyr.  We also note that the ratios $N_{\rm SSFS}/N_{\rm VNS}$ and $N_{\rm SS}/N_{\rm SSFS}$ in Table \ref{table:SFSS/VNS-Ratios}  do not appreciably change between the two epochs.  Moreover, although there is no restriction on the change in SFR for individual galaxies within each $M_*$ column, a direct comparison of the metallicities of galaxies in individual SFR/$M_*$ bins over a wide range of SFR and $M_*$ is still meaningful in telling us how the metallicities of the entire set have evolved in the 0.5 Gyr epoch covered.  


\begin{table}[t!]
\caption{Mid-z to Low-z Metallicity $Z_x$ Changes in Bins of $\geq 50$ Small-Radius SSFS Galaxies.}
\vskip0pt\vskip-6pt\vskip0pt
{\vskip-6pt}
\begin{center}
\begin{tabular}{lccccccccl} \hline\noalign{\vskip3pt}
$M_*(10^{10}M_{\odot})$&0.354&0.500&0.707&1.00&1.42&2.00&2.82&4.00&5.64\\
SFR($\frac{M_{\odot}}{{\rm yr}}$)&&&&&$10^3\Delta Z_x$&&&&\\
\noalign{\vskip3pt}\hline\noalign{\vskip3pt}
0.400    &    &    &    &    & & -0.05 & &   &    \\
0.566    & -0.006  & -0.031  & +0.060  & +0.09  & +0.04  & -0.03  & +0.09  &   &    \\
0.800    & +0.023  & +0.024  & +0.059  & +0.091  & +0.04  & -0.01  & +0.02  & +0.07  &   \\
1.13    & +0.056  & +0.046  & +0.030  & +0.061 & +0.10  & +0.04  & +0.02  & +0.05  &   \\
1.60    &    & +0.004  & +0.017  & +0.031  & +0.04  & +0.01  & +0.01  & +0.03  & +0.02  \\
2.26    &    &    & +0.065  & +0.043  & +0.01  & -0.01  & 0.00  & +0.02  & +0.02  \\
3.20    &    &    &    &    & +0.05  & 0.00  & +0.05 & -0.01  &  0.00  \\
\\\noalign{\vskip3pt}\hline
\end{tabular}
\end{center}
\label{table:metallicity-evolution}
\end{table}

We readily note that the quality of the medium-redshift data are rather lower than those obtained in the low-redshift range for samples with small radii.  The standard deviation of the metallicities cited at medium redshifts is similar, around 0.14 to $0.22\times 10^{-3}$, and the number of galaxies per bin peaks at 310, a factor of $\sim 2.5$ lower than among low-redshift  populations.  The standard error of the mean correspondingly increases by a factor of $\sim 1.6$, as is also visually apparent from the somewhat erratic jumps in metallicity from bin to bin, down $M_*$ columns in Figure \ref{table:SSFS-metallicity-Med-z} and the more deliberate progression down comparable columns in Figure \ref{table:SSFS-metallicity-Low-z}.  Between them, these factors lead to random errors of order the change in metallicity in some of the entries in Table \ref{table:metallicity-evolution} recording potential metallicity changes over an interval of $\sim 0.5$ Gyr.

Despite such concerns, Table \ref{table:metallicity-evolution} clearly indicates at most a distinctly low change in metallicity in small-radius galaxies observed at respective epochs 1.58 and 1.1 Gyr in the past.  For the 6592 galaxies represented in this tabulation of potential differences the mean fractional increase in metallicity over this interval, in galaxies of comparable size, appears to be as low as 3.4\% per galaxy, with a standard deviation of  $\sim 3.8\%$ over the 0.5 Gyr span separating the medium and low-redshift epochs.  Moreover, as discussed in more detail later in this section, part of such an apparent  metallicity increase would in any case be expected because at higher redshifts larger portions of a galaxy are encompassed within the SDSS fibers' fields of view.

This result has two implications:  

First, the mean metallicity evolution in the formation of comparable galaxies, over this 0.5 Gyr epoch, was at most of the order of a few percent, judged by these SDSS data, and presumably less than about one percent from one star-forming epoch, lasting of the order of a hundred million years, to the next onset of star formation.  

Second, throughout this 0.5 Gyr epoch, star formation rates in galaxies of identical mass were solidly anchored to identical metallicities.  If we take into account that evolved giant stars, as well as newly formed massive stars exploding as supernovae will have continually been generating metals throughout this epoch, while pristine hydrogen-helium mixtures may also have been funneled into these galaxies, it is striking that such a stringent equilibrium could have been maintained.  It is as though galaxies that had attained a given mass only 1.1 Gyr ago had undergone identical metallicity histories as had similar populations 0.5 Gyr earlier, despite an apparently ongoing increase in the metallicity of intragalactic gases throughout this epoch.  

Our finding of indeterminate metallicity evolution in our sample of SDSS galaxies is in apparent agreement also with the determination by \citet{Petr2012} of what appeared to be, at most, marginal increases in metallicities of low-mass galaxies, and then only in galaxies at the very centers of massive galaxy clusters, where intragalactic metallicities were especially high.  

\begin{figure}[h!]
	\centering
\includegraphics[height=1\textwidth,angle=90,trim=3.5cm 1cm .75cm 1.5cm, clip=true]{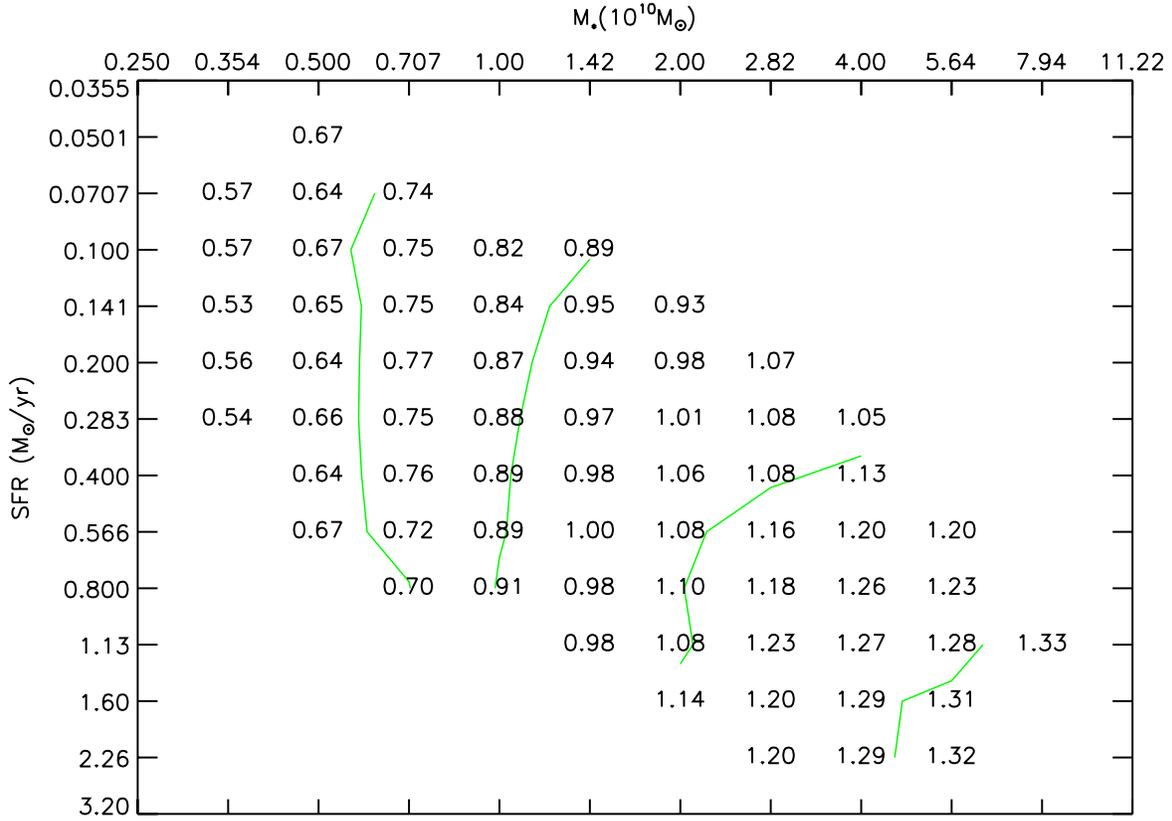}	
\caption{Metallicities (in units of 10$^{-3}$ (O/H)) of Low-z Medium-Radius SSFS Galaxies in Bins of $\geq 50$. Contours represent intervals of 0.2 with the bottom contour at 0.7.}
\label{table:Medium-radiusMetallicity}
\end{figure}

One other informative feature shown in Figure \ref{table:Medium-radiusMetallicity} is that low-redshift galaxies of medium radius, at all SFRs, exhibit on average roughly 5\% lower metallicities $Z_x$ than small-radius galaxies of identical masses $M_*$.  \citet{Elli2008} had already noted this effect many years ago; we present our result here primarily because our sample is more constrained in redshift and therefore more germane to our particular choice of sample and epoch. 

In Section \ref{section:parameters} we will note the critical role that Type II SNe play in enriching galaxies with oxygen.  In galaxies of identical mass, supernovae exploding at larger radii are more likely to permanently eject their metal-rich ejecta into extragalactic space because a galaxy's gravitational attraction is weaker at large radii and the viscous drag of lower-density ambient gases tends to be lower.   These galaxies'  metallicities, as judged by their retention of oxygen, may thus be expected to drop.  

\section{Rates of Galaxy Mass and Metallicity Evolution}\label{section:rates}

It is easy to see that, even at the highest SFRs $\sim 2.26 M_{\odot}$ yr$^{-1}$ in the lowest galaxy mass range $M_* = 3.54\times 10^9 M_{\odot}$ of Figure \ref{table:SSFS-metallicity-Med-z}, the stellar masses of these low-mass galaxies  can only increase by an amount $\Delta M_*\sim 10^9 M_{\odot}$ in the 0.5 Gyr interval over which galaxies listed in Figure \ref{table:SSFS-metallicity-Med-z} could potentially evolve into those listed in Figure \ref{table:SSFS-metallicity-Low-z} as they evolved from medium to low redshift.  This is not sufficient to move significant numbers of galaxies from one $M_*$ column of these tables into another.  Moreover, at the low redshifts discussed here, the SDSS finds hardly any galaxies accumulating mass through mergers of galaxies or major stellar aggregates.  We may, therefore, safely conclude that galaxies which were in any given $M_*$ column of Figure \ref {table:SSFS-metallicity-Med-z} at the beginning of the 0.5 Gyr period, will normally have continued to evolve into galaxies in the same $M_*$ column in Figure \ref{table:SSFS-metallicity-Low-z}.

This last point may need emphasis because the concept of ``downsizing" --- the fact that the population of small galaxies observed at low redshifts is much larger than the population of small galaxies seen at higher redshifts --- is sometimes interpreted to imply that ``the build-up of smaller galaxies (has occurred) at later epochs" \citep{Wong2012}.  Although the lower population observed at earlier epochs is readily explained as a selection effect, it is important to actually document that Figures \ref{table:SSFS-metallicity-Low-z} and \ref {table:SSFS-metallicity-Med-z} do not just represent a comparison of two distinct sets of galaxies of comparable mass as they appeared at different epochs.  Rather we are viewing a set of galaxies that, as a group, maintained a close-to-constant mass as they evolved over an epoch spanning 0.5 Gyr, and therefore occupy essentially identical $M_*$ columns in Figures \ref{table:SSFS-metallicity-Low-z} and \ref {table:SSFS-metallicity-Med-z}.

This consideration does not constrain galaxies in these two Tables from moving up or down within their $M_*$ columns over this 0.5 Gyr interval, as  star-forming rates periodically increased or decrease.  Nor would it constrain these galaxies from losing their SS designation and  to become SSFS galaxies or ceasing star formation entirely and becoming non-star-forming members of the VNS set.  Likewise, VNS or SSFS galaxies at medium redshift could have become SS galaxies over the interval of 0.5 Gyr
By and large, such trades in status cancel each other but, in any case, as Table \ref{table:SFSS/VNS-Ratios} shows, low-mass galaxies,  $M_* \lesssim 7\times 10^{9}M_{\odot}$, remain SS or SSFS galaxies throughout more than 90\% of their lives, and may thus be thought of as representing galaxies evolving directly from medium redshift, and thus inclusion in Figure \ref{table:SSFS-metallicity-Med-z}, to low redshift and inclusion in Figure \ref{table:SSFS-metallicity-Low-z}.

The extent to which metallicities might be expected to change over periods of order 0.5 Gyr, however, is less clear. Theoretical estimates by \citet{Krum2009} indicate that star formation efficiencies are  as low as  $\sim 1\%$, meaning that the portion of a galaxy's gaseous mass that forms stars during collapse is only $\sim 1\%$.  Taken at face value, this would imply that an episode of star formation leaves both the metallicity and the total mass of a galaxy's interstellar gas virtually unchanged.  Successive episodes of star formation in a galaxy should thus exhibit fairly constant metallicities, although slight increases could be expected as newly formed stars in successive episodes eject heavy metals, and as contributions to the galaxy's metallicity through mass loss from evolved stars  accumulates.  These increases may, however, be cancelled by the infall of pristine extragalactic gas or through the ejection of metal-enriched gases in supernova explosion of the most massive stars formed.

If metallicity then varies little from one star-forming epoch to another, should we expect star formation rates to similarly remain constant?  The close connection of SFR and metallicity exhibited in Figures \ref{table:SSFS-metallicity-Low-z} and \ref{table:SSFS-metallicity-Med-z} may suggest this;  but the metallicity values cited in the two tables have been averaged over more than 50 galaxies per bin, drastically reducing the standard error of the mean. In contrast, as noted in Section \ref{section:metallicity-evolution}, the standard deviation in the metallicities of individual galaxies is of order $0.2\times 10^{-3}$, i.e., at a level of $\sim 20\%$, meaning that the spread in metallicities of individual galaxies having identical star formation rates can range over an entire $M_*$ column.  Correspondingly, keeping a galaxy's metallicity fairly constant will not prevent abrupt changes in SFR.  Galaxies could thus easily wander between SS, SSFS, and VNS states, despite maintaining a constant metallicity.  Metallicity may thus not be a significant factor in determining SFRs, despite their apparent close correlation seen in Figures \ref{table:SSFS-metallicity-Med-z} and \ref{table:SSFS-metallicity-Low-z}.

Successive star forming episodes in a given cloud  may, nevertheless, give rise to fairly constant star formation rates.  If the star forming efficiencies of clouds remain as low as $\epsilon \sim1\%$, we may expect successions of star-forming episodes in which only small portions of a cloud collapse with star-formation rates varying in some proportion to fluctuations in $\epsilon$.  If these fluctuations then are of order unity, i.e., $\Delta \epsilon/\epsilon\sim 1$, fluctuations in SFRs would be similarly limited to $\Delta$SFR/SFR$\sim 1$.

All these considerations, however, require a more quantitative approach to which we turn next.  For this, it is best to deal directly with observable quantities, the most immediately useful among which we list in Table \ref{table:parameters}.

\begin{table}[t]
\caption{Observable Parameters of Selected Sources (SS), Strongly Star Forming Sources (SSFS) \& Valid Normal Sources (VNS), as well as Derived Parameters, and their Provenance}
\vskip0pt\vskip-6pt\vskip0pt
{\vskip-6pt}
\begin{center}
\begin{tabular}{lll} \hline\noalign{\vskip3pt}
Symbol&Parameter&Provenance\\
\noalign{\vskip3pt}\hline\noalign{\vskip3pt}
&\underline{OBSERVED PARAMETERS}&\\$M_*$&Galaxy Stellar Mass&From stellar luminosities\\
$N_{{\rm VNS}}|_{M_*}$&Number of $M_*$ VNS Galaxies&All SDSS VNS $M_*$ galaxies\\
$N_{{\rm SSFS}}|_{M_*}$&Number of $M_*$ SSFS Galaxies&All SDSS SSFS $M_*$ galaxies\\
$N_{{\rm SS}}|_{M_*}$&Number of $M_*$ SS Galaxies&All SDSS SS $M_*$ galaxies\\
SFR$(x)$&Star formation rate in ${M_{\odot}}$ yr$^{-1}$ &From H-$\alpha$ and dust opacity studies\\
$N_{{\rm SS}}(x)|_{ M_*}$&SS Population with SFR($x$)$|_{M_*}$&Number of SS $M_*$ galaxies, with SFR$(x)$\\
$N_s(x)|_{M_*}$&Number of galaxies with SFR$(x)|_{M_*}$&Number of $M_*$ galaxies, with SFR$(x)$\\
$Z_x$&Oxygen Abundances [O/H]&Spectra of galaxy H\sc ii \rm regions \\
&\underline{DERIVED VALUES}&\\
$\tau_s(x)/\tau_g$&Fractional SFR($x$) Population&From [$N_{{\rm SS}}(x)/N_{{\rm SSFS}}(x)][N_{{\rm SSFS}}/N_{{\rm VNS}}]$\\
$Z_0|_{M_*} $& $M_*$ Galaxy Peak Metallicity&From peak observed metallicities\\
$\alpha$&Low-mass Stellar Mass-Loss Rate&From stellar evolution studies\\
\noalign{\vskip3pt}\hline
\end{tabular}
\end{center}
\label{table:parameters}
\end{table}

\section{Star Forming and Fallow Episodes}

As pointed out in Section \ref{section:rates}, listing a given galaxy as a Valid Normal Source (VNS), a Strongly Star Forming Source (SSFS), or a Selected Source (SS), can only be a temporary designation.  An ongoing episode of abundant star formation may clearly identify a particular galaxy as belonging to the SS group, but as star formation wanes and the galaxy's H-$\alpha$ emission weakens it may recede into the SSFS range or cease star formation altogether and be assigned VNS status.  Renewed star formation may then restore the galaxy's SSFS or SS ranking.  Individual galaxies meander through these various phases as star-forming and fallow episodes succeed each other in as yet unpredictable sequences.

The duration of individual episodes will be determined by factors such as the IMF of the stars that are formed,  by the supply of gas to a galaxy's interstellar medium, and other factors.  But the mean evolutionary history of the galaxies can already be partially deciphered with the help of Table \ref{table:SFSS/VNS-Ratios}. For galaxies in each mass range $M_*$ the ratios $N_{{\rm SS}}/N_{{\rm SSFS}}$ and $N_{{\rm SSFS}}/N_{{\rm VNS}}$ tell us the fraction of the time these galaxies spend in different phases.  Thus we read that small-radius, low-redshift galaxies with masses in the range $M_* = 5.64\times 10^{10} M_{\odot}$ spend roughly 33\% of their lives as SSFS galaxies, and 67\% of their lives in a fallow state devoid of significant star formation but still meeting the more general VNS classification criteria.  Among all the VNS galaxies of this same mass and radius a typical galaxy would spend only a fraction ($N_{{\rm SS}}/N_{{\rm SSFS}}$)($N_{{\rm SSFS}}/N_{{\rm VNS}}$) $= 0.94\times 0.23 \sim 0.22$, i.e., 22\% of its life in the SS state.  For medium redshift galaxies in the same mass range the corresponding fraction would be $\sim 15\%$.  Quite generally, a comparison of the last two rows, respectively of the low-redshift and medium-redshift sections of Table \ref{table:SFSS/VNS-Ratios}, indicate that, within the fluctuations among ratios for galaxies of similar mass $M_*$, the fraction of their lives that galaxies spent actively forming stars remained virtually identical over the $\sim 0.5$ Gyr between the epochs at which galaxies in the two redshift ranges are being observed. 

We will be speaking, below, about typical cycles of duration $\tau_g$ in the lives of VNS galaxies, within which galaxies of a given mass $M_*$ forming stars at a rate SFR$(x)$ spend a period $\tau_{\rm SS}$ as SS galaxies, a period $\tau_{\rm SSFS}$ strongly forming stars, and only a period $\tau_{\rm VNS} - \tau_{\rm SSFS}$ in a fallow phase largely devoid of star formation. Table \ref{table:SFSS/VNS-Ratios} permits us to identify the relative lengths of such periods.  We can, for example, identify the fraction of the time $\tau_s(x)/\tau_g$ an $M_*$ galaxy spends over the eons forming stars at a rate SFR$(x)$, where $x$ represents the number of stars formed per year. Here the subscript `s' is meant to denote any star-forming galaxy, whether it be an SS or an SSFS galaxy.  However, we need to take care in defining just what we have in mind in discussing this fraction. 

We showed in Section \ref{section:rates} that the stellar masses of galaxies $M_*$ have not significantly changed between redshift ranges $z = 0.07$ to 0.10, or 0.10 to 0.15; but the rates at which individual $M_*$ galaxies were forming stars may have ranged widely throughout this time.   If we assume all the SDSS galaxies to have had such flexible histories, we may, for example, state that the fraction of its life a galaxy has spent as an SS galaxy forming stars at a rate SFR$(x_i)$, as contrasted to the time it has spent forming stars at any and all rates SFR$(x_j)$, is just the ratio in which the respective star forming galaxies are observed in the SDSS. 
\begin{equation}
\frac{\tau_{\rm SS}(x_i)}{\tau_{\rm SS}}\Bigg|_{M_*} = \frac{N_{\rm SS}(x_i)}{\sum_j N_{\rm SS}(x_j)}\Bigg|_{M_*}\equiv \frac{N_{\rm SS}(x_i)}{N_{{\rm SS}}}\Bigg|_{M_*}\ , 
\label{eq:ZxmassA}
\end{equation}
where $N_{\rm SS}(x_j)$ is the population of SS galaxies in the mass range $M_*$ listed in the SDSS as forming stars at a rate SFR$(x_j)$, $N_{\rm SS}(x_i)$ is the corresponding population of SS galaxies forming stars at a specific rate SFR($x_i$), and $\tau_{{\rm SS}}$ is the total time that SS galaxies of stellar mass $M_*$ spend in the Selected Source star-forming phase, during a cycle of duration $\tau_g$. 

On the other hand, if we are interested in the fraction of an $M_*$ galaxy's life spent as a Selected Source forming stars at a specified rate SFR$(x_i)$, rather than being in any alternative VNS state of mass $M_*$, we may obtain this ratio as
\begin{equation}
[\tau_{{\rm SS}}(x)/\tau_g]\Big|_{M_*} \equiv [\tau_{{\rm SS}}(x)/\tau_{{\rm VNS}}]\Big|_{M_*}   = [N_{{\rm SS}}(x)/N_{\rm VNS}]\Big|_{M_*}\ .
\label{eq:ZxmassB}
\end{equation}
where $N_{VNS}|_{M_*}$ is the number of $M_*$ galaxies in any VNS state, whether star-forming or fallow.

If a galaxy can abruptly change from being an SS galaxy producing stars at an arbitrary rate SFR$(x)$ to becoming fallow, without first passing through some arbitrary SSFS phase, or if quite generally any SSFS galaxy similarly forming stars at an arbitrary rate can likewise transit directly into a fallow state, this probability can be expressed as 
\begin{equation}
1 - \frac{\tau_s}{\tau_g}\Bigg|_{M_*} = 1 - \frac{\tau_s}{\tau_{{\rm VNS}}}\Bigg|_{M_*} = \frac{N_{\rm VNS} - N_s}{N_{\rm VNS}}\Bigg|_{M_*}\ ,
\label{eq:ZxmassE}
\end{equation}
where $N_s$ and $\tau_s$ can refer to either SS or SSFS galaxies. 
If we ask what fraction of the time an $M_*$ galaxy spends in any other phase than  forming stars at a rate SFR$(x)$ regardless of whether it does this in state SS or SSFS, this ratio reduces to
\begin{equation}
1 - \tau_s(x)/\tau_g\bigg|_{M_*} = 1 - \tau_s(x)/\tau_{{\rm VNS}}\bigg|_{M_*} =1 -  N_{{\rm s}}(x)/N_{{\rm VNS}}\bigg|_{M_*}\ ,
\label{eq:ZxmassF}
\end{equation}

We may, however, be more interested in the fraction of its life the galaxy spends in either a fully active SS, or at least a  partially active SSFS state, rather than in a fallow state forming few or no stars at all. This last-named state is represented by the ratio
\begin{equation}
\frac{\tau_s}{\tau_g}\Bigg|_{M_*} =  \frac{\tau_s}{\tau_{{\rm VNS}}}\Bigg|_{M_*}  =\frac{\sum_j N_s(x_j)}{N_{\rm VNS}}\Bigg|_{M_*}\equiv \frac{N_s}{N_{{\rm VNS}}}\Bigg|_{M_*}\ .
\label{eq:ZxmassC}
\end{equation}
 
We can also identify the fraction of the time a galaxy of stellar mass $M_*$ forms stars at a rate SFR$(x)$ while in an SS rather than in the more general SSFS state, and the relation of this ratio to the fraction of the time galaxies generally spend in such an SSFS state.
\begin{equation}
[\tau_{{\rm SS}}(x)/\tau_{{\rm SSFS}}(x)][\tau_{{\rm SSFS}}/\tau_{{\rm VNS}}]\bigg|_{M_*} = [N_{{\rm SS}}(x)/N_{{\rm SSFS}}(x)][N_{{\rm SSFS}}/N_{{\rm VNS}}]\bigg|_{M_*}\ .
\label{eq:ZxmassD}
\end{equation}



In this expression, the second term in square brackets on the right can be considered the conditional probability that a VNS galaxy is forming stars at a rate that qualifies it to be considered an SSFS galaxy rather than a galaxy essentially lying fallow.  The first term in square brackets on the right is the probability that an SSFS galaxy with mass $M_*$ forming stars at a rate SFR$(x)$ also meets the criteria to be ranked an SS galaxy.    $N_{\rm SS}(x)$, and $N_{\rm SSFS}(x)$, respectively, are the populations of galaxies at some particular galaxy mass $M_*$ forming stars at a rate SFR$(x)$, 
and the ratio $N_{{\rm SSFS}}/N_{{\rm VNS}}$ for $M_*$ galaxies is given in Table \ref{table:SFSS/VNS-Ratios}. 

Equations  (\ref{eq:ZxmassE}), (\ref{eq:ZxmassC}), and (\ref{eq:ZxmassD}) should not be interpreted as meaning that all SSFS galaxies spend equal fractions of their lives in a fallow state, independent of star forming rate SFR$(x)$.  This cannot be entirely correct, if nothing else, because galaxies transitioning into or out of a fallow state presumably pass through a low SFR state.    But, although the assumption may thus not be correct universally, Table \ref{table:SFSS/VNS-Ratios} shows the ratio $N_{{\rm SSFS}}/N_{{\rm VNS}}$ to be high over a considerable range of masses $M_*$ in our tables, suggesting that this assumption may not make much of a difference, at least not in low-mass $M_*$ ranges, because so small a fraction of these galaxies ever enters a fallow state.

Some of the ratios entering equations (\ref{eq:ZxmassE}) to (\ref{eq:ZxmassD}) are presented in Figure \ref{table:SSFS-Populations} and Table \ref{table:SFSS/VNS-Ratios}.   
We will be making use of such ratios in later sections and will need to take care in defining each particular choice.  As we do this, however, we will need to keep in mind that all these estimates of relative times spent in particular states hinge on the assumption that SSFS and SS galaxies move more or less randomly between various states SFR$(x_i)$ and SFR$(x_j)$, and that the relative populations found in these states are determined by some as yet unspecified circumstances.

\section{An Equilibrium Model of Star-Formation}\label{section:model}

Table \ref{table:parameters}, lists a number of observable parameters that the SDSS and other direct astronomical investigations provide.  

Our thrust will be to seek a relationship between four primary observable quantities, (i) the star formation rate SFR$(x)$ indicating the cumulative mass of stars formed per year; (ii) the metallicity $Z_x$ of the H$_{\rm II}$ regions within which the more massive stars have formed; (iii) stellar mass $M_*$ of the galaxy in which the stars are forming;  and (iv) the various galaxy population densities, such as $N_s(x)|_{M_*}$ sorted by the masses and star formation rates of the galaxies. One of the most significant contributions the SDSS provides is the link between SFR$(x)$ and the metallicity of the gas $Z_x$ in H$_{\rm II}$ regions enveloping newly formed stars.  This generally represents the metallicity of the gas from which the newborn stars originally formed, as distinct from a metallicity characterizing the ejecta from ambient massive stars in the assumed initial mass function that undergo supernova explosions in the first ten to twenty million years after the onset of star formation.

The SDSS galaxies divide rather starkly into two populations.  The lower mass galaxies in each of our redshift and radius subsets generally are active in star formation throughout, with a significant majority of them consistently and almost continually engaged in star formation as Table \ref{table:SFSS/VNS-Ratios} makes clear.  At the highest mass end, in contrast, relatively few of the galaxies are H-$\alpha$ emitters or engaged in forming stars. 

This leads us to include two further symbols in Table \ref{table:parameters}. The first, $Z_0|_{M_*}$ is the metallicity of the gas in a galaxy of stellar mass $M_*$ devoid of infall.  This can differ for galaxies of different masses.  Any pristine hydrogen falling into the galaxy dilutes this metallicity. For low-mass galaxies, the metallicity $Z_0|_{M_*}$ can be estimated, as shown in Paper I, by extrapolating the values of $Z_x$ in any given $M_*$ column back to zero star forming rates,  thus to the metallicity of a galaxy that may have lain fallow.  For more massive galaxies, which often are red, appearing to house solely older low-mass stars and devoid of star formation, the procedure is somewhat different. These galaxies exhibit exceptionally high metallicities $Z_x$ that do not vary significantly with SFR.  For these galaxies we can call their observed metallicity $Z_0$, which then also corresponds to their value of $Z_0|_{M_*}$. 

A recent paper by \citet{Sing2013} provides insight on the metallicities of these high mass galaxies.  The authors convincingly document that H-$\alpha$ emission in these galaxies cannot be attributed to star formation, but rather appears to be produced by ionizing radiation from post-AGB stars --- stars older than $\sim 1$ Gyr undergoing a brief but very hot and highly energetic phase following passage through the asymptotic giant branch of the Hertzsprung-Russell diagram.  

\citet{Sing2013} identify these galaxies as the --- now apparently misnamed --- low-ionization nuclear emission line region (LINER) galaxies. Their analysis of 48 LINERs shows the emission from these galaxies to be distributed more widely than merely the nuclear regions, and exhibits none of the steep rise in H-$\alpha$ luminosity that the earlier assumption of nuclear excitation had predicted.  The conclusion to be drawn from their study is that, contrary to the indications of \citet{Mann2010}, \citet{Lara2010}, \citet{Bris2012}, and the label SFR($M_{\odot}$/yr) heading the first column in each of the Figures \ref{table:SSFS-metallicity-Low-z} to \ref{table:Medium-radiusMetallicity}, the H-$\alpha$ emission should not be attributed unambiguously to star formation in galaxies more massive than $\sim 2 \times 10^{10} M_{\odot}$. Rather, the ionization in these galaxies is likely to be due, at least partially, to highly evolved or post-AGB stars. 

In their Figure 1, \citet{Kauf2003a} had identified many of the SDSS galaxies as LINERs as well, but had not reached the conclusion that their H$_{\rm II}$ emission was produced by ultraviolet radiation from post AGB stars.    

The observed H-$\alpha$ luminosity generated by such stars today is thus a delayed echo from a specific star-forming episode, $\sim 1$ Gyr earlier, its luminosity proportional to the luminosity of that earlier star-forming epoch. The observed metallicity of the ionized regions of high-mass galaxies should thus correspond to the metallicity of the reservoir of interstellar mass generated over time through mass-loss from these post-AGB and lower mass field stars.  As we will see in Sections \ref{section:enrichment} and \ref{section:parameters}, below,  mass loss from post-AGB stars tends to be nitrogen rich; lower mass stars generally are oxygen poor. A mixture of these appears to correspond well to the ratio of oxygen-to-nitrogen abundances observed in these galaxies.  Seen from this perspective the labeling of the first column of Figures \ref{table:SSFS-metallicity-Low-z} to \ref{table:Medium-radiusMetallicity} is not entirely misleading if we recognize that, for the post-AGB stars, the star-forming epoch referred to may have occurred 1 Gyr earlier and its magnitude would have had to be on some appropriately corrected scale --- the correction factor being of order unity.  This much can be concluded e.g. from \citet{Ilbe2013} who point out, e.g. in their Figures 10 and 14, that neither star formation rates, nor galaxy mass functions significantly changed between our redshifts $z = 0.07$ and 0.15 --- epochs separated by roughly 1 Gyr.
  
A related parameter $\alpha$ provides a measure of the mass of gas returned to the interstellar medium through mass loss from low-mass evolved stars.  Mass loss from these stars may be assumed to be proportional to a galaxy's stellar mass $M_*$ and lead to an integrated mass loss rate $\dot M = \alpha M_*$.  However,  low-mass stars do not contribute to oxygen enrichment; instead, they burn oxygen in their interior.  Their main contribution to processes discussed below is the hydrogen and helium they return to the interstellar medium as they evolve and lose mass.

\begin{figure}[h!]
	\centering
\includegraphics[width=15cm,trim=.25cm .25cm 0cm 12.5cm, clip=true]{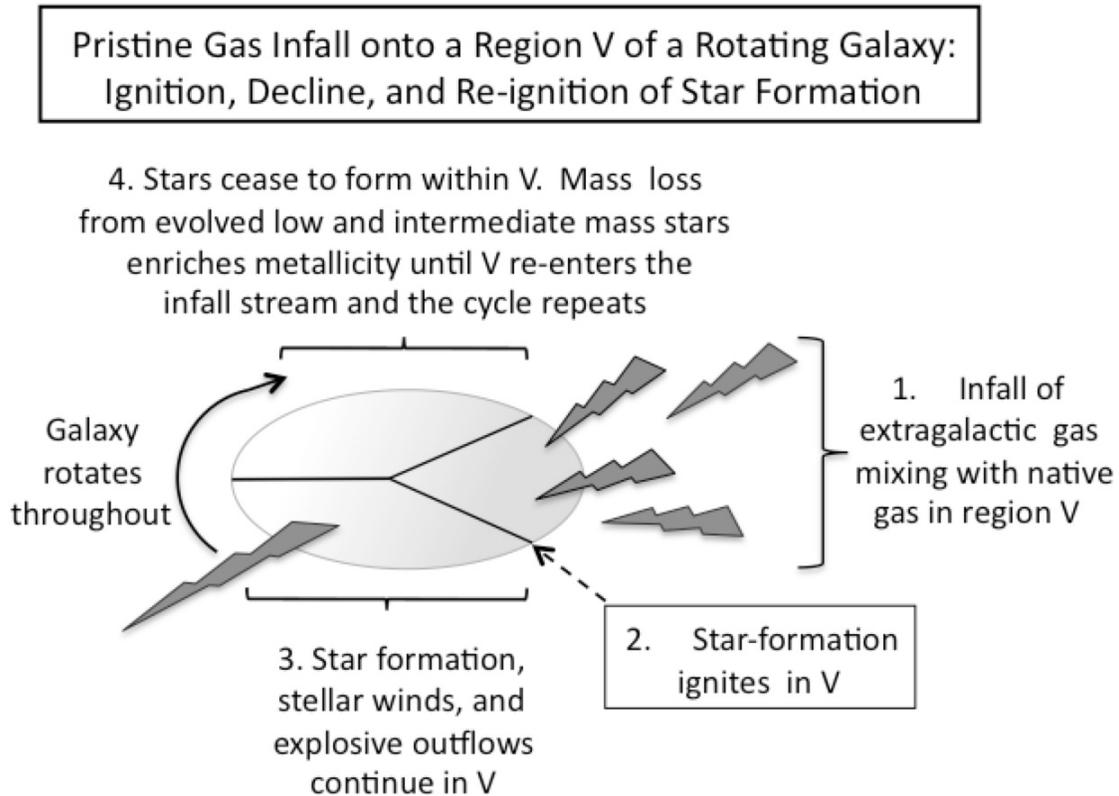}	
\caption{An Equilibrium View of Star Formation.  To an observer within a bounded galactic region $V$, star formation appears episodic.  Infall of extragalactic gas  (1) accumulates in $V$. When infall stops, star formation is triggered (2). Star formation then continues for an interval (3) until potential termination (4). The region then remains fallow until volume $V$ experiences further infall, possibly as the galaxy's rotation one again exposes $V$ to a persistent infall stream.  In contrast to the observer in $V$, an external observer exemplified by the SSDS sees the galaxy forming stars at a constant rate throughout.  No episodic changes are apparent. There is only a day side to the galaxy where infall triggers star formation and a night side where star formation fades but is revived as the next day dawns in the galaxy's rotation.}
\label{fig:SFGalaxy}
\end{figure}

The star-forming model of a galaxy we will be examining quantitatively is sketched in Figure \ref{fig:SFGalaxy}.   The strong correlation between high star formation rates and low metallicity in low-mass galaxies seen in Figures  \ref{table:SSFS-metallicity-Low-z} and  \ref{table:SSFS-metallicity-Med-z} has been widely interpreted as a causal connection between infall of pristine extragalactic gas and an onset of star formation.  By``pristine" we will for now consider gas of negligibly low metallicity. Starting in Section \ref{section:enrichment}, however, we will recognize the necessity of taking a low but significant infall metallicity into account. 

Star formation under the portrayed conditions is often considered to be episodic because the observed star formation appears localized.   Figure \ref{fig:SFGalaxy} indicates how localized star-forming patches may nevertheless be consistent with global star-forming equilibrium.  To an observer within region $V$, whose view of the entire galaxy is blocked by local interstellar dust, the duration of infall and subsequent star formation may appear quite short.   But an external observer viewing the entire galaxy will see star formation proceeding indefinitely, as long as the infall from extragalactic space continues.   This is the view of the galaxy the SDSS provides.  From this perspective the epoch over which star formation $\tau_s$ endures will equal the length of the epoch of infall $\tau_i$.  This is most readily seen for galaxies that continue to form stars steadily with few interruptions.  For these, the last phases of star formation will not be able to commence until the last phase of infall has subsided.  The star formation phase will always lag the infall phase with some delay, but if this delay is short compared to the duration of the infall epoch we will generally be able to set $\tau_s\sim \tau_i$.  For low-mass galaxies with $M_*\lesssim10^{10} M_{\odot}$, which appear to be forming stars almost full time, the validity of this approximation is particularly clear.

Periodically, infall from outside a galaxy may also cease altogether.  A galaxy's complete star forming cycle may then include extended fallow periods of duration $\tau_f$, so that the duration $\tau_g$ of a galaxy's complete star-forming cycle that includes such fallow periods, may significantly exceed the galaxy's rotation period $\tau_r$.

The infall of baryonic matter may be funneled into the galaxy along a filament of the cosmic extragalactic web.  After some time, for various reasons, the infall may cease, and a fallow period devoid of star formation sets in until infall may resume and a new episode of star formation begins.  

Each episode of star formation produces a family of massive stars that eventually explode and convey some of their metal-enriched ejecta into the extragalactic medium.  A balance between mass inflow and outflow, star formation, and the production, consumption, and ejection of heavy elements is therefore required if the unvarying metallicity of low-mass, small-radius galaxies documented in Section \ref{section:metallicity-evolution} is to be explained.

A star formation rate ${\rm SFR}(x)$ prevailing in region $V$ for time interval $\tau_s$ forms a mass of young stars given by 
\begin{equation}
{\rm SFR}(x) \tau_s(x) = \epsilon\times[\dot M_i\tau_i +\dot M_g\tau_g]\Big|_V \quad {\rm with}\quad \epsilon\equiv \epsilon(Z_x, M_*, \langle r\rangle, \langle z\rangle, \Pi)\ ,
\label{eq:Zxmass2}
\end{equation}
where the star-forming efficiency $\epsilon$ --- defined as the fractional mass of a collapsing gas cloud transformed into stellar mass --- may be a function of the  gas cloud's metallicity $Z_x$, the galaxy's stellar mass $M_*$, the mean galaxy radius $\langle r\rangle$ of the $M_*$ galaxies considered,  mean redshift $\langle z\rangle$ within our radius-redshift ranges, and potentially other parameters $\Pi$. $\dot M_i$ is the infall rate of pristine extragalactic gas averaged over the duration of infall $\tau_i$. $\dot M_g$ is the corresponding production rate of native gas, averaged over $\tau_g$, the duration of the entire star forming cycle in $V$.  $\dot M_g$ will generally have contributions from mass-loss by low-mass red giants or intermediate-mass evolved stars from earlier epochs of star formation whose most massive stars in their IMF will have long since exploded and passed away. Mass ejected by such stars may have been exploded out of the galaxy altogether, or may have fallen back into the galaxy, contributing  to $\dot M_g$, the net mass of ejecta retained in the galaxy.  In Section \ref{section:enrichment} we will consider each of these alternatives separately.

Infall of pristine extragalactic gas into galaxies, while inferred from existing data, has not yet been directly or reliably detected.  An alternative quantitative expression for the star formation rate based entirely on observable parameters may therefore have greater practical value.   

Integrated over a star-forming cycle of duration $\tau_g$ the mass of heavy elements consumed in star formation in equilibrium with infall at zero infall metallicity, $Z_i = 0$, is
\begin{equation}
{\rm SFR}(x) Z_x \tau_s(x)\Big|_V =\epsilon \tau_gZ_g\dot M_g\Big|_V\ ,
\quad 
\label{eq:Zxmass3}
\end{equation}
where $\tau_s(x)$ is the duration of the star-forming epoch that generates stars at a rate SFR$(x)$ and the expression on the right is the accumulation of gas with metallicity $Z_g$ in volume $V$ --- the integrated rate at which the galaxy's  stellar population returns mass.  Here the interval $\tau_g$ begins with the cessation of an earlier period of star formation in $V$ and ends once infall into $V$ has terminated and the newly formed stars begin to evolve.   

Contributions to the replenishment rate of heavy elements $Z\dot M_g$ provided by various classes of stars have been assessed from direct observations.  Preliminary assessments of the relative contributions from these varied sources were made by \citet{Oppe2008}  [O\&D(2008)] who concluded that low-mass AGB stars could not be the primary contributors to the oxygen abundances of most galaxies --- oxygen being the prime measure of metallicity, $Z$ used by \citet{Mann2010}, Paper I, and in the present paper.  Rather, the oxygen abundances in the gas from which stars and their ambient H$_{\rm II}$ regions form must have been contributed mainly by massive young stars because their ejecta exhibit a higher ratio of oxygen to nitrogen than expected from mass loss from lower-mass stars.

For galaxies having identical redshifts $z$, radii $r$, stellar masses $M_*$, and volumes $V$ within which stars are being formed, equation (\ref{eq:Zxmass2}) can be taken to be a statement of one form of Schmidt's law:  Integrated over time, star formation is proportional to the integrated mass of gas in a region $V$.  Equation (\ref{eq:Zxmass3}) expresses conservation of metallicity, in a form independent of infall rate $\dot M_i$ or duration $\tau_i$, as long as the metallicity of infalling material may be considered negligible.  This is useful since neither $\dot M_i$ nor $\tau_i$ have been directly observed, while most of the other parameters in equation (\ref{eq:Zxmass3}) are gradually emerging from observations.  

Equation (\ref{eq:Zxmass3}) may suggest that, for a fixed mass of stars, SFR$(x) \tau_s(x)$, formed from a fixed mass of native gas $\dot M_g\tau_g$, $\epsilon$ should be directly proportional to $(Z_x/Z_g)$ and independent of the mass of pristine gas falling into the galaxy.  This may, however, be misleading if the mass of infalling gas $\dot M_i \tau_i$ directly determines the amount of star formation a given amount of native gas can support. 

Nevertheless, since the metallicities derived from SDSS data refer primarily to oxygen abundances, any correspondence between the efficiency $\epsilon$ and metallicity $Z(x)/Z_g$ could be attributed to enhanced cooling rates by silicate dust, by gas enriched in carbon monoxide CO, or by water vapor H$_2$O and would thus be plausible because each of these oxygen-bearing interstellar constituents is recognized as a primary cooling agent in different phases of molecular cloud collapse. 

At this stage it may be useful to summarize specific features of the model to which we will be returning in the remainder of this paper: 

1) We portray star formation as a steadily ongoing process.  Although a cursory glance at data may give the impression that star formation is punctuated by outbursts, viewed over longer periods or on a larger scale, a more regulated sequence of events appears to emerge.   As new stars form, the more massive die out more rapidly but are replaced by younger equally massive stars as the region in which star formation is active gradually moves to neighboring locations.  

2) Infall of matter from extragalactic space, in some as yet unspecified fashion, actively promotes star formation.  Matter does rain in on galaxies episodically, as evinced by the fraction of VNS galaxies nearly devoid of star formation; but some stellar mass ranges $M_*$ show nearly continuous star formation, where $(N_{{\rm SSFS}}/N_{{\rm VNS}})\sim 1$, in Table \ref{table:SFSS/VNS-Ratios}, and for them these fallow phases can be rare.  

3)   A galaxy houses a population of low-mass evolved stars proportional to the galaxy's total stellar mass $M_*$.  Their steady mass loss enriches the metallicity of interstellar gas albeit with matter of relatively low oxygen content.  Metal-enriched material with higher oxygen content is ejected by massive young stars, but an as yet undetermined fraction of this material escapes the galaxy propelled by explosive velocities ranging in the many hundreds of kilometers per second.  These  two sources determine the metallicity $Z_g$ of gas native to the galaxy.  In high-mass, predominantly-red galaxies, the contributions of low- and possibly intermediate-mass AGB stars may be heightened, judged by their higher [N/O] ratios.  In low-mass galaxies, massive young stars are dominant contributors to the abundance of oxygen. 

4) The metallicity of gas observed in star forming regions, $Z_x$,  arises from the balance of three quantities --- (i) an ongoing accumulation of mass loss from low-and intermediate-mass evolved stars, (ii) the explosive injection --- and partial retention --- of metal-enriched supernova ejecta and (iii) the infall of pristine, largely metal-free gas.

5) The extent to which a specific metallicity level controls the star formation rate is unclear, but the strong correlation of mean metallicity  $Z_x$ and SFR$(x)$ in bins of $\geq 50$ galaxies, independent of redshift, is unmistakable in Figures \ref{table:SSFS-metallicity-Low-z} and \ref{table:SSFS-metallicity-Med-z}, and Table \ref{table:metallicity-evolution}. 

6)  Despite phases of abundant star formation, the stellar masses $M_*$ of SDSS galaxies have remained virtually constant over the 0.5 Gyr between redshift $z = 0.1$ and 0.07, and the metallicity of galaxies of comparable mass and SFRs has remained essentially unchanged as well, within differences of the order of the observational uncertainties ($\sim 3\%$). 
   
7) Although SFRs and metallicities, clearly are functions of galaxy stellar mass $M_*$ and radius $r$, as originally noted by \citet{Elli2008}, the decline in metallicity resulting from accumulating infall rarely drops below $Z_x \sim 0.7\times 10^{-3}$,  roughly half the peak metallicity of native gas $Z_0|_{M_*}$ before star formation triggers, and further dilution ends. 

8) A prominent feature of this model is the cyclic, repetitive nature of its star-forming episodes under long-term constant infall.  With so many features of low-redshift galaxies remaining close to invariant, successive star forming episodes within a galaxy appear likely to also differ little --- a feature consistent with this model.  Repetitive phases, however, could also be expected on independent grounds, namely, because star-forming efficiencies are believed to be as low as 1\% \citep{Krum2009}, suggesting that star formation serially starts, quenches, revives, and quenches again, in a series of bursts successively consuming small, roughly identical portions of a star-forming cloud.

\section{Metal Enrichment from Massive, Intermediate- and Low-Mass Stars}
\label{section:enrichment}

In a fallow period $\tau_g$ preceding or overlapping onset of star formation in some volume $V$ of a galaxy, the metallicity of gas in $V$ can incrementally change.  This increase or decline of interstellar metallicity is determined by four processes leading to five potential outcomes: (i) enrichment or dilution through infall of extragalactic matter, (ii) enrichment or dilution through mass loss from evolved low-mass stars, (iii) enrichment through delayed ejection of matter from intermediate-mass stars, or (iv) return of mass initially exploded from massive stars but either promptly admixed with interstellar matter or falling back into the parent galaxy with some delay before being admixed, and (v) a fraction of this explosively ejected matter entirely escaping the parent galaxy. This last outcome enriches extragalactic space but does not raise the metallicity of the parent galaxy's interstellar gases. We will call the four classes of process: Class I, infall; Class A, actively persisting return of mass lost by low-mass stars; Class D, delayed mass loss generally from more massive asymptotic giant branch stars; and Class E, explosive ejection leading to either of the last two outcomes (iv) or (v).

(i) Class I: Extragalactic mass falling into a galaxy at a rate $\dot M_i$ averaged over a time interval $\tau_i$ may contribute its own metallicity $Z_i$ at a rate:  
\begin{equation}
(\tau_i/\tau_g) \dot M_i Z_i\ \sim (N_{{\rm SSFS}}/N_{{\rm VNS}})\dot M_i Z_i\ ,
\label{EnrichmentI}
\end{equation}
where $\dot M_i$ is the mass infall rate of extragalactic matter of metallicity $Z_i$, and the approximation on the right assumes, as in Section \ref{section:model}, that the duration of infall, $\tau_i$, roughly equals the duration of star formation, $\tau_s$, i.e., $\tau_i\sim\tau_s$.  In this section we will concentrate our attention on SSFS galaxies, since it is now clear that galaxies move seamlessly between SS and SSFS phases, dependent largely on the extent of their SFRs albeit, as previously pointed out, consistently retaining nearly constant stellar mass $M_*$.

(ii) Class A: Mass loss of gas by evolved low-mass, $\lesssim 2 M_{\odot}$ giant branch stars, at a rate $\dot M_g$ and metallicity $Z_A$, adds metals at a rate $\dot M_gZ_A|_V$.  Here $\dot M_g$ is the production rate of gas at metallicity $Z_A$  in the portion of an $M_*$ galaxy contained in the star-forming volume $V$.  Neither the fractional population of old stars within volume $V$, nor the metallicity of their ejected material is expected to greatly vary, so that the contributed mass of gas should be proportional, through a constant $\alpha$, to the galaxy's stellar mass $M_*$ enclosed in volume $V$.  We thus obtain a metal enrichment rate
\begin{equation}
\dot M_gZ_A\bigg|_V = \alpha Z_A\,M_*\bigg|_V\ .
\label{eq:EnrichmentA}
\end{equation}
This rate is expected to prevail both during active and fallow epochs, and is often a greatly delayed influx from earlier-formed generations of low-mass stars undergoing mass-loss toward the end of their lives.  When the volume $V$ encompasses the entire galaxy this mass loss is simply proportional to $M_*$.  

For native gas production through the Class A process an equilibrium metallicity $Z$ for the mix of relatively pristine infalling gas with metallicity $Z_i$ and native interstellar matter having metallicity $Z_g = Z_A$, can be expressed as: 
\begin{equation}
Z = \frac{(\dot M_i \tau_i Z_i+ \dot M_g\tau_g Z_A)}{(\dot M_i\tau_i + \dot M_g\tau_g)}\Bigg|_V\ .\quad   {\rm If}\ Z_i \ll Z_A\quad {\rm then}\quad Z _x= \frac{\dot M_g\tau_g Z_A}{(\dot M_i\tau_i + \dot M_g\tau_g)}\Bigg|_V\ .  
\label{eq:Zxmass1}
\end{equation}
For aged low-mass stellar populations, $Z_A$ is not expected to vary with galaxy stellar mass $M_*$.  If the infall stops, i.e., $\dot M_i = 0$, $Z = Z_A$, and the metallicity is determined by stellar mass loss.  If the infalling gas has zero metallicity, $Z_i = 0$, galaxies that have accumulated an infall of $\dot M_i\tau_i$ equal to an accumulation of native gas, $\dot M_g\tau_g$, i.e., $\dot M_i\tau_i=\dot M_g\tau_g$, should exhibit a metallicity $Z$ equal to half the native metallicity, $Z = \frac{1}{2}Z_A$ due to the mass loss from low-mass evolved stars.  Such low metallicities are observed, if at all, solely in low-mass galaxies exhibiting high SFRs. 

The metallicity $Z_A$ of a galaxy's native gas before it becomes diluted by infalling gas is most readily obtained for high-mass galaxies.  These undergo significant fallow periods, as judged by their relative population densities $N_{\rm VNS} > N_{\rm SSFS}$, and exhibit consistently high $Z$ values, $\sim Z_A$, even at the higher SFRs for which reliable data are available.  In contrast, in low-mass galaxies even low SFRs and correspondingly low infall rates appear to appreciatively dilute the galaxies' gas content.

Care must be taken in interpreting equation (\ref{eq:Zxmass1}).  The metallicities $Z_i$, $Z_A$ and $Z$ all refer to the abundance by number, in our case of oxygen to hydrogen, [O/H], respectively in the infalling gas, the native gas in a galaxy, and the mixture of the two. This is also the sense in which Mannucci et al (2010) used the term {\it metallicity}, as did our Paper I. Other authors have used that same nomenclature to denote the fractional abundance by mass of a particular element to the total mass of all elements, meaning that account has to be taken not only of oxygen and hydrogen, as we are doing, but also of helium and other elemental abundances.  We will return to this distinction in Section \ref{section:parameters}. 

(iii) Class D:   Matter expelled in delayed mass loss from intermediate-mass evolved stars is generally of much higher metallicity than that of the H$_{\rm II}$ regions  listed in Figures \ref{table:SSFS-metallicity-Low-z}, \ref{table:SSFS-metallicity-Med-z}, and \ref{table:Medium-radiusMetallicity}.

Intermediate-mass stars having masses $2\lesssim M\lesssim 10M_{\odot}$ and spectral type ranging from B8 to B0 while on the main sequence  are not sufficiently massive to collapse and explode as supernovae, but are nevertheless sufficiently massive to evolve from the main sequence in less than 1 Gyr \citep{Scha1992}.  On becoming giants their cores contract to form white dwarfs or neutron stars, even as they shed their outer layers to enrich the metallicity of the interstellar medium from which a new generation of stars will form. 

For the enrichment of oxygen O\&D(2008) draw a distinction between  AGB stars in the mass range $4\leq M\leq8 M_{\odot}$ and those in the range 2 to 4 $M_{\odot}$.  Only the former generate an abundance of oxygen, evolve rapidly, and thus their mass loss is significant even at high redshifts.  The lower-mass stars take longer to evolve to the AGB phase, and the earliest-formed among them at high redshifts will not have evolved sufficiently until much later and much lower redshifts before they begin to shed mass.  Mass loss from the higher-mass stars can thus be observed even at high redshift, whereas the loss from lower-mass stars can only be expected at later times --- meaning lower redshifts.  

The metallicity enrichment by these intermediate-mass stars then has to be understood to have originated at some earlier epoch whose star formation rate SFR(x) may not have been identical to a galaxy's current star formation rate --- although, as discussed in Section \ref{section:model}, the differences could be minor if the galaxies undergo successions of star-forming events that differ little from one another. 

Note that Class A and D may be thought of as two sides of one and the same coin.  Viewed from a Class A perspective, delayed return of processed matter through stellar mass loss goes on continually independent of whether or not a galaxy is forming stars throughout.  The rate of return must then be prorated across all time.  In contrast, process D concentrates on the production epoch, and considers delayed return to be proportional to the length of time star formation was active at an earlier epoch.  The mass loss is then attributed to the fraction of intervals $N_{\rm SSFS}/N_{\rm VNS}$ during which star formation had been active earlier. An accompanying difference between processes A and D is that, seen from the vantage point of process D, a specific star formation rate SFR$(x)$ accounts for a particular yield at some designated later epoch.  In contrast, process A assumes a designated initial mass function, takes into account the fraction of all stars $M_*$ that have formed within a galaxy, and asserts that the steady state return of processed material averaged over time, is simply determined by the IMF and $M_*$.  No particular star forming epoch with a specified SFR$(x)$ needs to be called on to account for this steady state yield. The ratio $N_{\rm SSFS}/N_{\rm VNS}$, plays no role in processes considered Class A, but does enter in the context of Class D: 

\begin{equation}
\dot M_D(x)Z_D\Bigg|_{M_*} \equiv (Z_D\epsilon_D)\langle{\rm SFR}(x)\rangle\Biggl(\frac{N_{\rm SSFS}}{N_{\rm VNS}}\Biggr)\Bigg|_{M_*} .
\label{eq:EnrichmentD} 
\end{equation}
Here, $\epsilon_D$ is the efficiency of forming stars in the mass range $\sim 4$ to 8 $M_{\odot}$  leading to delayed oxygen enrichment.  Because we primarily consider galaxies clustered in bins of more than 50 galaxies having identical star formation rates, $\langle{\rm SFR}(x)\rangle$ may be considered to be the mean delayed star-forming rate these galaxies experienced at earlier epochs. For stars in the 4 to 8 $M_{\odot}$ mass range this earlier epoch tends to be $\lesssim 1$ Gyr earlier and thus should be rather similar to current conditions, as discussed in Section \ref{section:rates}.  We may therefor take $Z_D$ to be a metallicity weighted appropriately to the mean star formation rate in the $\sim 4$ to 8 $M_{\odot}$ stellar mass range.

Class E:  Supernovae produce a prompt enrichment of a galaxy's interstellar medium through the fraction of the explosively ejected metals retained in the galaxy  and its rapid mixing with the ambient interstellar medium.  This is most likely the fate of ejecta expelled roughly parallel to a galaxy's disk, where the ejecta's high initial velocities are most rapidly damped through viscous interaction with ambient interstellar gas. 

Supernovae, however, also produce matter that may rise to considerable heights in the parent galaxy's gravitational potential before eventually falling back.  Although the fallback is delayed, this hardly matters in  episodic star formation. If the material falling back does not admix with interstellar matter during a current star forming episode it will contribute its metals at some subsequent epoch, so that only the summed contributions averaged over time need taking into account.  Averaged over time the fall-back components may thus be included in the prompt supernova return.  We will call the cumulative fraction of the ejected mass of Type II SNe ultimately retained in their parent galaxies, $f_{E,R}$.  This is a function of both $M_*$ and SFR$(x)$.  The rate of Class E enrichment then is 
\begin{equation}
\dot M_E(x)Z_E\Bigg|_{M_*} = \frac{N_{{\rm SSFS}}{\rm SFR}(x) (Z_E\epsilon_Ef_{E,R})}{N_{\rm VNS}}\Bigg|_{M_*}\ ,
\label{eq:EnrichmentE}
\end{equation}
where $\epsilon_E$ is the fractional mass of stars in the IMF undergoing Type II SNe explosions, and $Z_E$ is the metallicity of the ejecta. 

Considering all four of the processes, we can formally write, in analogy to the metallicity derived in equation (\ref{eq:Zxmass1}),
\begin{equation}
Z_x\Bigg|_{M_*} = \frac{\alpha M_*Z_A\tau_g  +[\dot M_D(x)Z_D + \dot M_E(x)Z_E+ \dot M_iZ_i]\tau_s}{\alpha M_*\tau_g +  [\dot M_D(x) + \dot M_E(x)+ \dot M_i]\tau_s}\Bigg|_{M_*}\ ,
\label{eq:OverallMetallicity0}
\end{equation}
where $\dot M_i$ is the instantaneous infall rate.  More explicitly,
\begin{equation}
Z_x\Bigg|_{M_*} = \frac{\alpha M_* Z_AN_{\rm VNS} + N_{\rm SSFS}[Z_{D}\epsilon_{D}\langle {\rm SFR}(x)\rangle  + {\rm SFR}(x) Z_E\epsilon_Ef_{E,R} + \dot M_iZ_i]}{\alpha M_*N_{{\rm VNS}} +  N_{{\rm SSFS}}[\langle {\rm SFR}(x)\rangle\epsilon_{D} +  {\rm SFR}(x)\epsilon_Ef_{E,R} + \dot M_i]}\Bigg|_{M_*}\ .
\label{eq:OverallMetallicity1}
\end{equation}
The fraction $\epsilon_E$ then encompasses only two types of supernova ejecta, a fraction $f_{E,R}$ involved in prompt return and the fraction $(1- f_{E,R})$ escaping a galaxy altogether. A number of authors have defined a hybrid delayed enrichment D due to matter initially ejected from, but later falling back into a galaxy, as in supernova explosion models discussed by \citet{Oppe2010}.  Mass ejected from these galaxies may then fall back with a delay of several hundred million years.  With suitable modifications such models could be subsumed in either of our Classes D or E. 

Inverting equation (\ref{eq:OverallMetallicity1}), we obtain the infall rate associated with a particular star formation rate SFR$(x)$ and enrichment rates $Z_A$, $Z_{D}$, and $Z_E$.   
\begin{equation}
\dot M_i(x)\Bigg|_{M_*} =\Biggl[ \alpha M_* \Biggl(\frac{Z_A-Z_x}{(Z_x-Z_i)}\Biggr)\frac{N_{{\rm VNS}}}{N_{{\rm SSFS}}} + \epsilon_D\Biggl(\frac{Z_{D}-Z_x}{(Z_x-Z_i)} \Biggr)\langle {\rm SFR}(x)\rangle + \epsilon_Ef_{E,R}\Biggl(\frac{Z_E -Z_x}{(Z_x-Z_i)}\Biggr){\rm SFR}(x) \Biggr]\Bigg|_{M_*}\ .
\label{eq:OverallInfallRate}
\end{equation} 

A consideration with all four processes, I, A, D, and E is that accumulating high-metallicity matter must somehow be prevented from leading to a rapid rise in metallicity over time, since we found no indication of enrichment, at least in small-radius galaxies, even over a period as extended as 0.5 Gyr.  Two factors may prevent rapid change over recent epochs.  First, continual infall of pristine material from extragalactic space inevitably dilutes accumulating higher-metallicity interstellar gas that may subsequently form new stars; and star formation preferentially incorporates enriched interstellar material, more adept at radiating away heat to form stars.  Between them, these two processes may regulate the metallicity to maintain it constant.

\section{The Physical Parameters Governing Star Formation and Metallicity}
\label{section:parameters}

In Paper I we demonstrated that, for SS galaxies, the infalling gas appeared to be pristine, with $Z_i = 0.125\times 10^{-3} \pm 0.125 \times 10^{-3}$, consistent with zero metallicity.   The low metallicity values we derived were based on our assumption that even the oxygen rich ejecta of Type II SNe, which contribute most to a galaxy's oxygen metallicity, would always remain within their parent galaxy, rather than escaping  to extragalactic space.    

Once this assumption is dropped, higher infall metallicities need to be considered.  A study of intracluster gases by \citet{Lovi2011} shows that oxygen abundances in the space between galaxies in clusters rich in metals declines rapidly with increasing distance from the centers of the five clusters they studied, A496, A2029, Centaurus, Hydra A, and S\'{e}rsic 159-03. The three most distant of these, all with redshifts still somewhat lower than the galaxies in our samples, exhibited lower metallicities than the two nearer clusters, Centaurus and A496.  At distances of 300 kpc from the cluster centers, the mean metallicities for Hydra A, A2029, and S\'{e}rsic 159-03, were of order 25\% solar, and apparently still declining, though with large error bars.  Metallicities this low correspond to $Z_i \sim 0.20\times 10^{-3}$ --- just under 50\% of the lowest metallicities in any of the galaxy bins in our Figures  \ref{table:SSFS-metallicity-Low-z}, and \ref{table:SSFS-metallicity-Med-z}, and about  15\% the metallicity of our most massive galaxies.

A large majority of our SDSS galaxies lie beyond these central cluster distances \citep{Hoyl2012}.  We may therefore expect the metallicity of their infalling gas to be no higher than $Z_i \sim 0.20\times 10^{-3}$ and quite possibly significantly lower.  As a first approximation, we will take the metallicity of the infalling gases to be, $Z_i \sim 0.20\times 10^{-3}$, until we may check this in Section \ref{section:Cross-check} below, or if improved observational data become available. Equations (\ref{eq:OverallMetallicity0}), (\ref{eq:OverallMetallicity1}) and (\ref{eq:OverallInfallRate}) will readily permit insertion of other $Z_i$ values as more accurate data accrue.

To see how well equation (\ref{eq:OverallInfallRate}) corresponds to a representative current theory, we may examine the paper by \citet{Oppe2008}.  Other theoretical approaches could be similarly checked, but O\&D(2008) are especially clear in detailing their assumptions, particularly in providing separate estimates of the oxygen production rates of Type I and Type II SNe and by AGB stars.  
\subsection{Prompt and Delayed Metal Yields}

To make best use of the approach of O\&D(2008) a correspondence between their terminology and that used by \citet{Mann2010}, Paper I, and the present paper may be helpful.  Table \ref{table:parameters2} defines some of the required symbols.

Restricting ourselves, for the moment, to Type II SNe ejecta, which contribute most of the oxygen to the metallicities $Z_x\equiv$ [O/H], with which we have dealt throughout the present paper, we need to recall that [O/H] has consistently represented the ratio of oxygen atoms to hydrogen atoms in the H$_{\rm II}$ regions in star-forming SDSS galaxies. 

\begin{table}[t!]
\caption{Symbols Referring to Infall, SFR Efficiencies, Metallicities, Yields, and Mass Loss}
\vskip0pt\vskip-6pt\vskip0pt
{\vskip-6pt}
\begin{center}
\begin{tabular}{ll} \hline\noalign{\vskip3pt}
Symbol&Parameters\\
\noalign{\vskip3pt}\hline\noalign{\vskip3pt}
$\epsilon_D$&Fraction of IMF Contributing to Delayed Mass Loss\\
$\epsilon_E$&Fraction of IMF Producing Type II SN Formation\\
$f_{E,R}$&Fraction of Type II SNe Ejecta Retained in a Galaxy\\

$\dot M_i$& Instantaneous Mass Infall Rate\\
$\dot M_{\ell}(x)$&Rate of Mass Ejection from a Galaxy through SFR$(x)$\\
$\dot M_{O,P}$, $\dot M_{O,D}$&Rate of Prompt, respectively, Delayed Oxygen Production\\
${\rm [O/H]}$&Ratio of Oxygen Atoms Relative to Hydrogen: See also $Z_x$\\
$\{{\rm O/M}\}$&Oxygen Mass as a Fraction of Total Mass of Stellar Ejecta\\
$Y_D$&Delayed Mass Yield of a Metal\\
$Y_P$&Prompt Mass Yield of a Metal\\
$Z_E\equiv Z_{O,P}$&Metallicity by ratio of atoms [O/H] of Type II SNe Ejecta\\
$Z_D$ &Delayed Stellar Mass Loss Metallicity by Ratio of Atoms [O/H]\\
$Z_{O,P}\equiv Z_E$&Promptly Produced Oxygen Metallicity by Ratio of Atoms [O/H]$_P$\\
$Z_{O,D}$&Delayed Oxygen Metallicity Produced by Ratio of Atoms [O/H]$_D$\\ 
$Z_i$& Metallicity of Infalling Gas by Ratio of Atoms [O/H]\\
$Z_x$&Oxygen Metallicity by Ratio of Atoms [O/H]\\
\noalign{\vskip3pt}\hline
\end{tabular}
\end{center}
\label{table:parameters2}
\end{table}

In the terminology of O\&D(2008) a distinction arises between two types of {\it yields}  a {\it prompt yield}, $Y_P$, contributed by Type II SNe explosions, and a {\it delayed yield}, $Y_D$, lost by less massive stars of the IMF, generally delivered 0.1 to 1 Gyr after the onset of star formation, by which time these stars have considerably evolved. 
Both yields represent the rate at which the mass of a given element, in our case mainly oxygen, is built up as a function of the star formation rate --- both these rates being measured in terms of mass per unit time.

Thus, the prompt yield of oxygen $Y_{O,P}$ is defined as the total mass per unit time of the promptly produced oxygen divided by the star formation rate:
\begin{equation}
Y_{O,P}\equiv \frac{\dot M_{O,P}}{{\rm SFR}} = \frac{\epsilon_E \langle f_{E,R}\rangle {\rm SFR}\{{\rm O/M}\}}{{\rm SFR}} \equiv \{{\rm O/M}\}\epsilon_E\langle f_{E,R}\rangle\ ,
\label{eq:O/M}
\end{equation}
where, in the second equality on the right, $\epsilon_E$ is the fraction of the IMF, by mass, giving rise to Type II supernova ejecta; the fraction $\langle f_{E,R}\rangle $ refers to the fraction of all Type II SNe ejecta retained in the galaxy, on average; and  \{O/M\} is the mass of oxygen as a fraction of the total mass of all these ejecta --- hydrogen, helium, and metals.  Some authors distinguish between supernova ejecta immediately mixed with ambient gas and retained in a galaxy, and material initially expelled to significant distances from the galaxy but ultimately falling back to be retained.  We will understand the retained fraction $f_{E,R}$ to comprise both these forms of ejecta.
To obtain the oxygen to hydrogen ratio $Z_x \equiv$ [O/H], by number of the respective atoms, we divide the mass of oxygen by its atomic number, 16, and divide again by  0.72 because the fraction of hydrogen, by mass, in a solar-abundance hydrogen-helium mix is only $\sim 72\%$, the rest being largely helium. So, finally, the mean metallicities averaged over the distribution of retention rates of supernova ejecta is 
\begin{equation}
\langle Z_{O,P}\rangle \equiv \langle Z_E\rangle \equiv [{\rm O/H}]|_P =\frac{Y_{O,P}}{(16\times 0.72)\epsilon_E\langle f_{E,R}\rangle} = \frac{0.0868Y_{O,P}}{\epsilon_E \langle f_{E,R}\rangle}\ , 
\label{eq:OP}
\end{equation}
where we make explicit that $Z_{O,P}$ refers to the Class E metallicity, $Z_E$ in equations (\ref{eq:OverallMetallicity0})  to (\ref{eq:OverallInfallRate}).  

O\&D(2008) define their delayed oxygen yield in somewhat different form --- as the fraction of oxygen, by mass, contained in the mass loss from evolved stars:
\begin{equation}
Y_{O,D}= \frac{\langle\dot M_{O,D}\rangle}{\epsilon_D\langle{\rm (SFR)}\rangle} = \frac{\epsilon_D\langle{\rm (SFR)}\rangle\{{\rm O/M}\}}{\epsilon_D\langle{\rm (SFR)}\rangle}= \{{\rm O/M}\}\ .
\end{equation}
Here $\epsilon_D$ is the ratio of mass loss by evolved stars to the mass of newly formed stars.
Dividing this, as before, to obtain the atomic ratio [O/H], we arrive at: 
\begin{equation}
Z_{O,D}\equiv Z_D \equiv [{\rm O/H}]_D = Y_{O,D}/(16\times 0.72)\ . 
\end{equation}

\subsection{Oxygen Metallicity}

The most massive stars in a stellar association's initial mass function  contribute oxygen both promptly on exploding as Type II SNe, and with a delay typically of order 0.1 to 1 Gyr as delayed mass loss from intermediate mass stars.  Using data provided by O\&D(2008) we can derive a probable mass loss to the intergalactic medium due to supernova ejection. In their Figure 6f, O\&D(2008) estimate the prompt oxygen yield in Type II supernovae.  This yield is the mass of prompt oxygen released by supernovae in a star-formation cycle, divided by the stellar mass formed in the cycle. The prompt yield, $Y_{O,P}\sim 1.5\times 10^{-2}$, can be considered the fractional mass of oxygen produced and promptly ejected in a cycle.  This is about 5 times the estimated yield of delayed oxygen produced to date, $Y_{O,D}\sim 3\times 10^{-3}$.  O\&D(2008) also estimate the cumulative delayed mass production by redshift $z\sim 0$ as $\epsilon_D\sim 38\%$ of the total mass of all stars formed.  This translates into a delayed oxygen production to date of $ M_{O,D}\sim 1.1 \times 10^{-3}$ of all stellar mass ever formed. 

The prompt and delayed production rates add up to a total oxygen production rate of $\dot M_{O,P} + \dot M_{O,D}\sim 1.6\times 10^{-2}$ of all stellar mass formed.  We may compare this to the solar oxygen abundance by mass, $M_{O,\odot}\sim 9.6\times 10^{-3}$ obtained from the estimates of \citet{Ande1989}.  Since both the prompt and delayed oxygen yields of O\&D(2008) have remained constant over the past few Gyr, this apparent anomaly needs an explanation.  However, it should not suggest that solar abundances imply a mean retention $\langle f_{E,R}\rangle\sim 0.6$ at the sun.  Rather, it appears to reflect a persistent problem with solar abundances, \citep{Haxt2013}.  Helioseismic data suggests that the abundance of heavy elements must be higher in the solar interior than observed on the solar surface and that the chemical composition of the sun is far from homogeneous throughout.

Returning to equation (\ref{eq:OP}), and inserting the O\&D(2008) prompt oxygen yield $Y_{O,P} = 1.5\times 10^{-2}$, we obtain
\begin{equation}
\epsilon_E\langle Z_Ef_{E,R}\rangle = 0.0868 Y_{O,P}\sim 1.3\times 10^{-3}\Biggl[\frac{Y_{O,P}}{(1.5\times 10^{-2})}\Biggr]\ .
\label{eq:Y_{OP}}
\end{equation}
 O\&D(2008) estimate that $\epsilon_E \sim20\%$ of the mass of stars distributed in an IMF undergoes supernova explosions.    

The delayed enrichment of oxygen provided by AGB stars, again referred to by O\&D(2008) as their {\it delayed yield}, is the rate of oxygen production divided by the rate of mass loss from evolved stars, i.e. $Y_{O,D} = \dot M_{O,D}/\epsilon_D {\rm (SFR)} \sim 3\times 10^{-3}$. This makes the fraction [O/H] by number of atoms, $Z_{O,D} = 0.0868 Y_{O,D} \sim 0.26\times 10^{-3}$, where we have similarly taken the fractional amount of ejected helium into account.

To summarize our adopted values from O\&D(2008):  For delayed release, $\epsilon_D = 0.38,\ Z_D = 0.26\times 10^{-3}$.  For prompt release from Type II SNe we set  $\epsilon_E=0.2$.  Our best estimate for the metallicity of infalling matter is $Z_i = 0.2\times 10^{-2}$.   We will derive values of $Z_E \equiv Z_{O,P}$ in Section \ref{section:Cross-check}.

\subsection{Conservation Relations}

For mass-loss from AGB stars we pursue the interpretation of O\&D(2008) and portray delayed enrichment as a process of type D subsuming both processes A and D defined in Section \ref{section:enrichment}.  With this simplification we can rewrite equation (\ref{eq:OverallInfallRate}) as :
\begin{equation}
\dot M_i(x)Z_i + \epsilon_DZ_D\langle{\rm SFR}(x)\rangle + \epsilon_EZ_Ef_{E,R}{\rm SFR}(x) = Z_x\biggl[\dot M_i(x) +\epsilon_D\langle{\rm SFR}(x)\rangle +\epsilon_Ef_{E,R}{\rm SFR}(x)\biggr] \ .
\label{eq:MetallicityConservation}
\end{equation}

Care needs to be taken in interpreting this equation.  It states that the metallicity of a region in which stars are forming is the mean metallicity of matter that had accumulated in the region before the new generation of stars began to form.  The ambient H$_{\rm II}$ regions exhibit a metallicity identical to that from which the stars formed.  

$M_{\ell}(x)$, the rate of mass loss from a galaxy through supernova expulsion will then be  
\begin{equation}
\dot M_{\ell}(x) = \epsilon_E (1- f_{E,R}) {\rm SFR}(x)\ ,
\label{eq:MassEjectionRate}
\end{equation}
where $f_{E,R}$ is generally a function of galaxy mass $M_*$ as well as the range of star formation rates assumed.

\section{A Cross-Check for Self-consistency}\label{section:Cross-check}

We need to verify that the sequence of entries in Section \ref{section:parameters} are mutually self-consistent.  The main uncertainty among the parameters we defined centers on $f_{E,R}(x)$ the fraction of supernova ejecta retained in the galaxy.   We need to evaluate this next. 

We noted earlier that the metallicity of galaxies at low redshift remains essentially constant, at least over periods of order 1 Gyr.  This means that the metallicity of infalling gas when admixed with gases retained from an immediately preceding cycle of star formation must equal a constant metallicity $Z_x$ derived from the weighted metallicities of infalling and retained gases.  This is just the relation given in equation (\ref{eq:MetallicityConservation}), which we can rewrite as
\begin{equation}
\dot M_i (x)\Bigg|_{M*} = \frac{\epsilon_E f_{E,R}(x)(Z_E-Z_x){\rm SFR}(x)}{(Z_x - Z_i)} - \frac{\epsilon_{D}(Z_x -Z_D)\langle{\rm SFR}(x)\rangle}{(Z_x-Z_i)}\Bigg|_{M*}\ .
\label{eq:MetallicityConservation'}
\end{equation}
In addition, in equilibrium, the fractional mass of gas in galaxies of a given mass $M_*$ should  remain relatively constant, rather than systematically increasing or decreasing over time.  The mass of infalling gas plus whatever ejected mass is retained in the galaxy and admixes with it should thus, on average, equal the mass of stars formed. 
\begin{equation}
\dot M_i (x) +  \epsilon_Ef_{E,R}(x){\rm SFR}_j(x) + \epsilon_D \langle {\rm SFR}(x)\rangle  = {\rm SFR}_{(j+1)}(x)\ .
\label{eq:MassBalance}
\end{equation}
Here, the left side of the equation sums the gas production rates available for star formation at the end of a star forming episode $j$.  The right side of the equation conserves this, converting it into star forming rates in an immediately succeeding episode $(j+1)$.  In Sections  \ref{section:model} and \ref{section:enrichment} we pointed out that the low star forming efficiencies expected from dynamic models make likely that successive episodes of star formation in a given galaxy or massive cloud may give rise to fairly similar rates of star formation differing from one episode to the next only by factors of order unity. This similarity would also hold if infall into a galaxy were relatively stable for extended periods, as suggested by the cyclic model described in Section \ref{section:model}.  If so, the quantities ${\rm SFR}_{j}(x)$, $\langle {\rm SFR}(x)\rangle$, and ${\rm SFR}_{(j+1}(x)$ should all be roughly of the same magnitude.  Averaging over 50 or more galaxies in a given bin such as those in Figure \ref{table:SSFS-metallicity-Low-z} should then enable us to approximate equation (\ref{eq:MassBalance}) by
\begin{equation}
\dot M_i(x) = \left[1 - (\epsilon_E f_{E,R}(x) + \epsilon_D)\right]{\rm SFR}(x)\ . 
\label{eq:MassBalance2}
\end{equation}
 
An equivalent way of stating this is to maintain that, in our equilibrium model: (i) the mass of infalling gas, plus (ii) the retained mass of gas from stellar mass loss in an immediately preceding cycle of star formation, should account for (iii) the mass of stars formed in the ensuing cycle.  Not specifically called out on the right side of (\ref{eq:MassBalance}) is the mass retained within stars once all their mass loss has ceased; nor does the right side include any additional mass ejected from the galaxy.  Neither of these contributes to the galaxy's retained gas content.

Combining equations (\ref{eq:MetallicityConservation'}) and (\ref{eq:MassBalance}) to eliminate $\dot M_i(x)/$SFR$(x)$, we obtain an expression for  $f_{E,R}(x)$, the fractional mass of stellar ejecta retained within the galaxy
\begin{equation}
f_{E,R} = \frac {(Z_x-Z_i) - \epsilon_D(Z_D-Z_i)}{\epsilon_E(Z_E-Z_i)}\  .
\label{eq:f}
\end{equation}

To estimate $Z_E$ we examine this expression in the high $M_*$ ranges where $Z_x$ reaches a maximum around $\sim 1.35 \times 10^{-3}$ at relatively high star formation rates. The close-to-invariant metallicities of these galaxies indicates that they are sufficiently massive to retain all Type II SNe ejecta, so that we may set $f_{E,R}= 1$.  Setting $Z_x = 1.35\times 10^{-3}$, constant over this high-$M_*$ star-forming population, and with $Z_i = 0.2\times 10^{-3}$ and $\epsilon_E = 0.2$ this leads to $Z_E = 5.85\times 10^{-3}$ .

With $Z_E$ thus defined, we can use equation (\ref{eq:f}) to evaluate $f_{E,R}(x)$ as function of $Z_x$,  
\begin{equation}
f_{E,R} = \frac{10^3 Z_x - 0.22)}{1.13}\ .
\label{eq:numeric}
\end{equation}
The ratio of infall to star-formation rate $\dot M_i(x)$/SFR$(x)$ is then  obtained from equation (\ref{eq:MassBalance}), with results shown in Figure \ref{table:SSFSInfall/SFR}.


\begin{figure}[h!]
	\centering
\includegraphics[height=1\textwidth,angle=90,trim=3.5cm 1cm .75cm 1.5cm, clip=true]{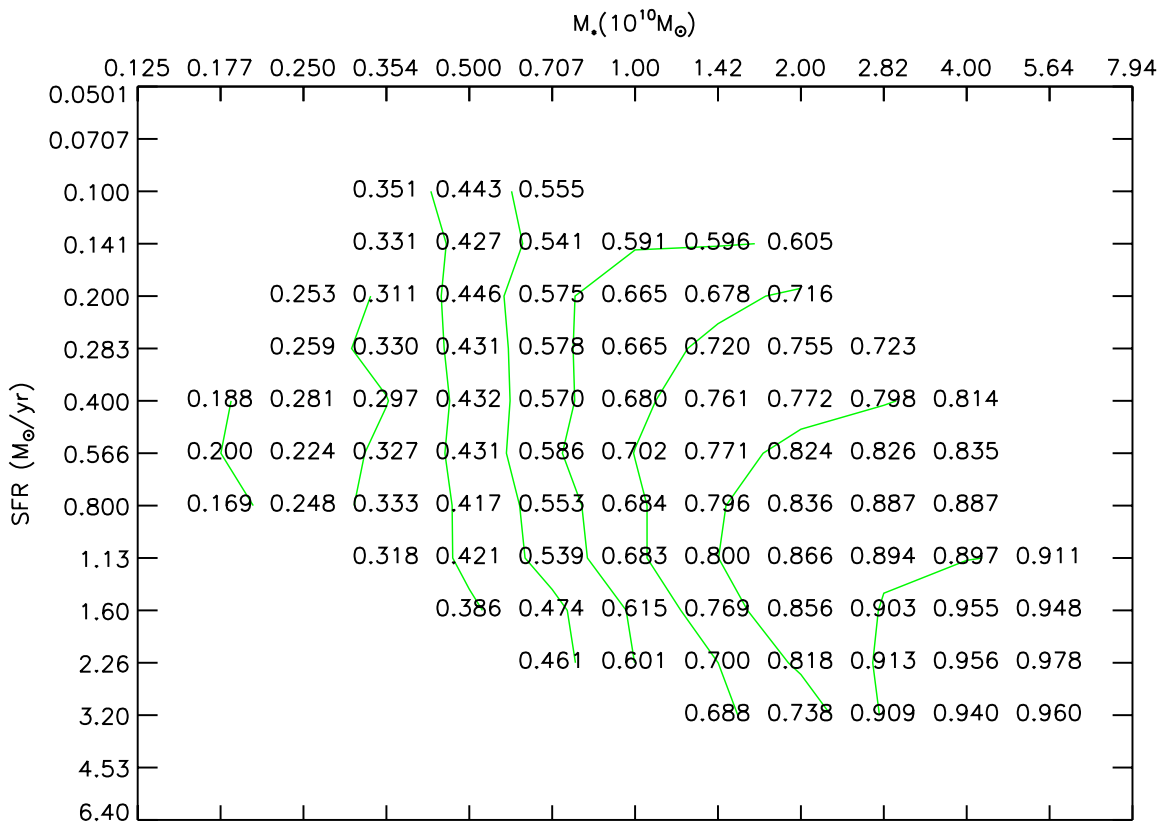}	
\caption{Retention Rates, $f_{E,R}(x)$, in bins of $\geq 50$ Low-z Small-Radius (SSFS) Galaxies. Contours represent intervals of 0.1 with the bottom contour at 0.2.}
\label{table:SSFSretention}
\end{figure}




We may still evaluate the mean value of $\langle f_{E,R}\rangle$ using our entries for SSFS galaxies in Figures \ref{table:SSFS-Populations} and \ref{table:SSFS-metallicity-Low-z}.  This is obtained using equation (\ref{eq:numeric}) and the mean oxygen abundance in galaxies, yielding:
\begin{equation}
\langle Z_x\rangle = \Biggl[\sum_{{\rm SFR}(x)}\sum_{M_*} N(x){\rm SFR}(x)Z_x\Bigg/ \sum_{{\rm SFR}(x)}\sum_{M_*} N(x){\rm SFR}(x)\Biggr] \sim 1.03\times 10^{-3}\ ,
\label{eq:retention1}
\end{equation}
averaged over galaxies populating bins of more than 50 galaxies.  Averaging over all galaxies, regardless of the number per bin, the mean metallicity drops negligibly to $\sim 1.02\times 10^{-3}$. 
The mean retention rate $\langle f_{E,R}\rangle$ of oxygen within galaxies then is
\begin{equation}
\langle f_{E,R}\rangle = \frac {(\langle Z_x\rangle -Z_i) - \epsilon_D(Z_D-Z_i)}{\epsilon_E(Z_E -Z_i)} = \frac{\langle Z_x\rangle -0.22\times 10^{-3}}{1.26\times 10^{-3}} \sim 0.644
\label{eq:retention2}
\end{equation}
evaluated for SSFS galaxies falling into bins containing $\geq 50$ galaxies; this again drops negligibly to $\sim 0.637$ when including all SSFS galaxies regardless of bin size.  


A point worth noting is that a significantly higher infall metallicity than the $Z_i \sim 0.2 \times 10^{-3}$ we have postulated, would require excessively high infall rates to dilute the native gas in low mass galaxies in Figure \ref{table:SSFS-metallicity-Low-z} down to observed metallicities as low as $Z_x=0.41 \times 10^{-3}$.  For the moment, therefore, a value of $Z_i \sim 0.2\times 10^{-3}$ seems reasonable.  Improved observations of infall metallicities should eventually also lead to improved estimates of mean retention in galaxies, $\langle f_{E,R}\rangle$, and mean ejection, $\langle (1 - f_{E,R})\rangle$.


\begin{figure}[h!]
	\centering
\includegraphics[height=1\textwidth,angle=90,trim=3.5cm 1cm .75cm 1.5cm, clip=true]{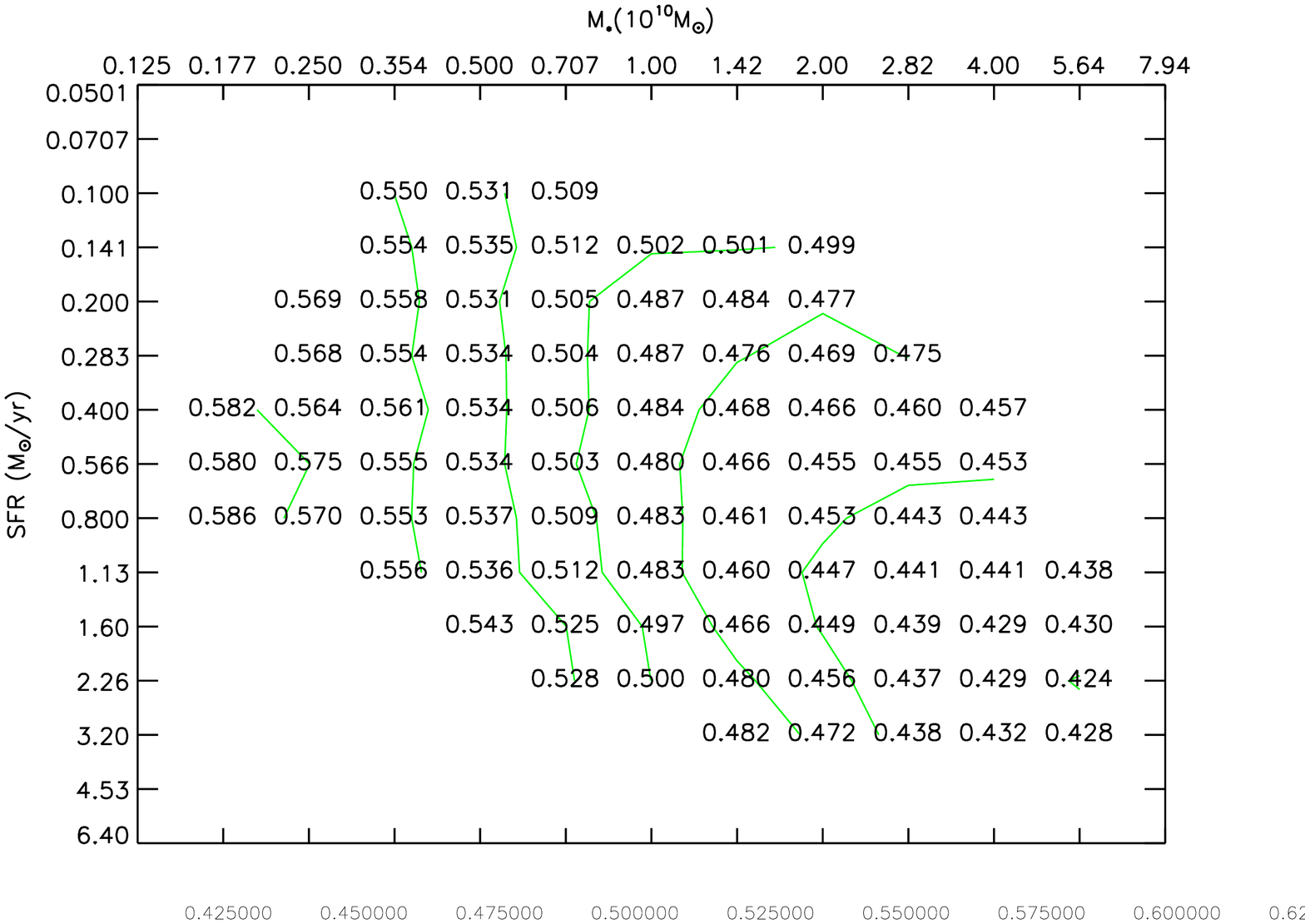}	
\caption{Specific Infall $\dot M_i/{\rm SFR}(x)$ in bins of $\geq 50$ Low-z Small-Radius (SSFS) Galaxies. Contours represent intervals of 0.025 with the bottom contour at 0.425.}
\label{table:SSFSInfall/SFR}
\end{figure}

\subsection{Injection of Metals into Extragalactic Space} 

A further question of interest regards the relative metallicities of infall and outflow.  Equilibrium between the metallicities of infall and outflow would be reached if the rates of infalling metals were to equal their outflow:
\begin{equation}
\Bigg\langle \frac{\dot M_i}{{\rm SFR}(x)}Z_i\Bigg\rangle \Longleftrightarrow \langle (1 - f_{E,R})\rangle \epsilon_EZ_E\ .
\label{eq:in/out}
\end{equation} 
Using values based on $Z_i = 0.2\times 10^{-3}$, we find that fractional metal outflow on the right side of expression (\ref{eq:in/out}) is roughly a factor of $\sim 5$ higher than the infall.  Simple estimates indicate that this outflow rate, nevertheless, is not sufficiently high to significantly increase the metallicity of the extragalactic medium over periods of order 1 Gyr in the low-redshift universe.

Another estimate, however, emerges from the work of O\&D(2008) discussed in Section \ref{section:parameters}.  It shows that the fractional mass of oxygen produced to date in all the stars ever formed should be of order $Y_{O,P} \sim 1.5\times 10^{-2}$. The corresponding metallicity averaged over all baryonic matter should accordingly be $[{\rm O/H}]\sim 0.0868Y_{O,P}(N_*/N_{XG})\sim 2\times 10^{-4}$, where $N_*$ and $N_{XG}$, respectively, are the number of cosmic baryons in stars and in extragalactic gases.  

This is close to our estimate of the metallicity of infalling gas; but, depending on the prevalence of intergalactic gases it might account for today's intergalactic metallicity only if the expulsion of gas from galaxies had consistently been of order unity, rather than our mean value $\langle (1 - f_{E,R})\rangle \sim 0.36$ apparently characterizing average galaxies today.   Current estimates of the ratio of baryons in intergalactic space to baryons in stars remain quite uncertain, ranging from roughly 5:1 to 1:1, respectively estimated for intracluster gases by \citet{Gonz2013} and for the warm/hot intergalactic medium, WHIM, \citet{Byko2008}.   At the high end it would be difficult to account for the observed metallicities of \citet{Lovi2011} even if high-redshift galaxies had been considerably less massive than today and their expulsion of oxygen-rich material could correspondingly have been much higher, roughly of the order of 80\% as observed in low mass galaxies today.

With this proviso, the self-consistency of our approach may at least come into rough agreement with still-uncertain observational constraints, particularly on the number density of baryons in extragalactic space.

\section{Discussion and Conclusion}\label{section:conclusion}

The thrust of this paper has been a theoretical investigation of the relation between galaxy stellar mass, metallicity, and star formation rate, noted by the observations of \citet{Mann2010} and \citet{Lara2010}, and the modeling of Dav\'{e} and Oppenheimer, 2008. To this end we have produced, and made available online, tables characterizing metallicity and sample size as a function of star formation rate and stellar mass for our full data set broken into subsets of redshift range and radial extent.

Our work has restricted itself to the redshift range $0.07 < z \leq 0.3$, with emphasis on galaxies in the range $0.07 < z \leq 0.15$.  Here the mass-metallicity-star-formation relation is well documented and the wealth of data provided by the Sloan Digital Sky Survey permits a thorough check.  At larger redshifts the validity of the relation is not as clear, although the work of  \citet{Sava2005} indicates that it may still apply at intermediate redshifts.  But it would not be surprising to find that it does not hold at high redshifts where galaxy mergers and starbursts were far more common than in the local universe. 

Our study has been largely based on two conservation principles.  The first is the conservation of gas or, more accurately stated, interstellar matter. The mass of stars formed plus the mass explosively ejected from a galaxy, must equal the mass of interstellar matter provided through infall from extragalactic space plus mass loss from evolved stars.  The second is conservation of metals.   Their appearance or disappearance at different times in the evolution of stars and galaxies needs to be accounted for at each stage.  

An advantage of using metallicity as a tracer is that infall and outflow estimates, which are difficult to make on dynamic grounds can at least be reasonably bounded by arguments based on metallicity, as we have tried to show.  As increasingly accurate data on stellar evolution and mass loss accrues, improved bounds on the interplay between the intragalactic medium and star formation should come within reach.  The use of other tracers, such as carbon, silicon, or iron, should ultimately provide added insight to that provided by oxygen alone.  Each of these tracers is produced or consumed in stars of different mass, and is injected into interstellar or extragalactic space at differing rates.  Tracing their different histories will thus provide distinct perspectives on how the interplay of stars, galaxies, and extragalactic gases has evolved over the eons. 

We are aware that improved techniques for assessing metallicities continue to be adopted, so that it is clear that an analysis of the kind we have pursued will need to be repeated as newer methods come into use, particularly as metallicity determinations are updated using far-infrared [O III] metallicity determinations. As pointed out in recent papers by \citet{Crox2013} and \citet{DeLo2014}, far-infrared abundances are considerable less dependent on the temperatures of H$_{\rm II}$ regions, and a composite picture based on a compendium of far-infrared spectral lines should soon lead to more accurate estimates of star formation rates.


The equilibrium model we have produced is based largely on observations, and is thus phenomenological.  A first consideration is that, at low redshifts, galaxy metallicities are observed to not significantly change over periods of order 1 Gyr. Galaxy evolution is further constrained by taking the gaseous component of galaxies to remain roughly constant during the low-redshift era, as distinct from inexorably increasing or decreasing over time.  These two restrictions lead to an equilibrium model that appears consistent with: (i) observations on the metallicity of gases in galaxy clusters, and (ii) consequently also the metallicity of gases falling into today's galaxies; (iii) the observed metallicities of ${\rm H_{II}}$ regions enveloping new stars, and (iv) their relation to the observed SFR; (v) the fraction of gases ejected from Type II SNe and their retention within, or ejection from, their parent galaxies; (vi) the relative metallicities of gases observed in small- and medium-radius galaxies; and a number of other observations. 

We have shown that the galaxy-mass-metallicity-star-formation relation can be traced to infall of extragalactic gas mixing with native gas from host galaxies to form stars of observed metallicities, the most massive of which eject oxygen into extragalactic space.  These processes can also account, very roughly, for the metallicity of extragalactic space, though the fraction of all baryons currently in extragalactic space will need to be more firmly established to check this.

Potentially the most consequential of our findings is that, on average, at low redshifts, extragalactic infall accounts for roughly half of the gas required for star formation. This ratio of infall to native gas is remarkably constant across galaxies with stellar masses ranging at least from $M* = 2 \times 10^9$ to $6\times 10^{10} M_{\odot}$.  This leads us to propose that star formation is initiated when extragalactic infall roughly doubles the mass of marginally stable interstellar clouds.  

The finding is remarkably robust, and virtually immune to different ways of assessing metallicities.  \citet{Yate2012} have examined the differences between metallicity estimates employed by \citet{Mann2010} and \citet{Trem2004}. These remain substantial, particularly in the low metallicity ranges, where differences of the order of 40\% are not unusual.  These are the ranges reflecting the most massive outflows, so that a more accurate accounting would be highly desirable.  Despite such differences, however, we have found that the $\dot M_i/{\rm (SFR)}$ ratios displayed in Figure \ref{table:SSFSInfall/SFR} remain remarkably stable.  Using data originally acquired using the methods of \citet{Mann2010} and stored on line as part of Paper I, we have carried out comparisons of $\dot M_i/{\rm (SFR)}$ ratios that turned out to be in surprisingly good agreement with values presented in Figure \ref{table:SSFSInfall/SFR}.  We find differences amounting to a systematic increase across the table of only $\sim 5-8\%,$ in all locations, but with identical trends ranging from similarly elevated ratios for galaxies of low mass to a lower ratio for high-mass galaxies.  That the $\dot M_i/{\rm (SFR)}$ ratio should be so insensitive to differences in the way metallicities are established is reassuring, but more consistent  ways of establishing metallicities remain a high priority.  The work of \citet{Crox2013} suggesting that metallicities based on far-infrared fine-structure transitions will yield superior results is thus welcome news.  Data already available in the Herschel archives may bring assurance that this resource will be of help. 

A question that will need to be more carefully studied, however, is the nature of the ${\rm H_{II}}$ regions in high-mass galaxies.  The analysis of \citet{WuYZ2013} of relative abundances of nitrogen to oxygen in galaxies of different masses makes clear that the oxygen observed in low-mass galaxies must largely have been produced in and ejected by, massive stars, whereas the oxygen observed in high-mass galaxies appears to be largely contributed through mass-loss by evolved intermediate-mass stars,  potentially with masses in the 2 to $4M_{\odot}$ range.  Stellar evolution models constructed  by \citet{Meyn2002} first indicated that AGB stars in this general mass range, particularly stars with high rotational velocities, should be expected to lose mass with exceptionally high [N/O] ratios.  If the ${\rm H_{II}}$ regions in high and low mass galaxies eventually are confirmed to have such different origins, the constancy of the $\dot M_i/{\rm (SFR)}$ ratio that we find across the entire mass range we examined may need further elucidation. 

The results we have derived do not depend on specific dynamic models, but dynamic models will eventually need to be tested against our findings on relations such as those between galaxy stellar mass $M_*$, star formation rates and metallicities, and the infall and outflow parameters derived from them.  Once sufficiently reliable metallicity tables are in hand various calculations will become possible.  Among others, we will be able to analyze rates at which metals are injected into the extragalactic medium by galaxies of different masses $M_*$ and radii, as a function of redshift. The SDSS already provides data for many of these calculations at low redshifts.   Our formalisms for dealing with such questions can thus be applied to a sizable range of investigations to at least provide simple but informative answers to questions that otherwise might require far more complex calculations. 

One factor still needing mention is that assumptions leading to equation (\ref{eq:numeric}) should be further tested. No galaxy may be totally immune to loss of interstellar matter and its embedded metals --- if not through supernova ejection directly into extragalactic space, then through a gradual but systematic drift of interstellar matter into the central portions of a galaxy where it may disappear into a supermassive black hole or be ejected into extragalactic space along jets powered by the black hole\ \citep{Chen2013,Rosa2013}.  Equation (\ref{eq:numeric}) should therefore be recognized as an interim estimate based on our current limited understanding.

Our model leads to a large number of predictions, particularly those based on abundances of other metals like carbon, iron or silicon, whose stellar yields are known.  Some of these predictions may be tested even now, while others may have to await the design of new observations.  In particular, the predicted mass losses from galaxies and their metallicities are currently becoming observable. Infall of gas and its metallicity may soon also be satisfactorily observed to test our model's merits.

\section*{Acknowledgment}

The present study was supported by NASA through subcontracts 1393112 and 1463766 the Jet Propulsion Laboratory awarded to Cornell University.

The National Radio Astronomy Observatory is a facility of the National Science Foundation operated under cooperative agreement by Associated Universities, Inc.

We are indebted to an anonymous referee, who suggested a number of improvements we implemented.

\bibliography{mybibliography.bib}

\begin{thebibliography}{43}
\expandafter\ifx\csname natexlab\endcsname\relax\def\natexlab#1{#1}\fi

\bibitem[{{Anders} \& {Grevesse}(1989)}]{Ande1989}
{Anders}, E., \& {Grevesse}, N. 1989, \gca, 53, 197

\bibitem[{{Brisbin} \& {Harwit}(2012)}]{Bris2012}
{Brisbin}, D., \& {Harwit}, M. 2012, \apj, 750, 142

\bibitem[{{Bykov} {et~al.}(2008){Bykov}, {Paerels}, \& {Petrosian}}]{Byko2008}
{Bykov}, A.~M., {Paerels}, F.~B.~S., \& {Petrosian}, V. 2008, \ssr, 134, 141

\bibitem[{{Cardelli} {et~al.}(1989){Cardelli}, {Clayton}, \&
  {Mathis}}]{Card1989}
{Cardelli}, J.~A., {Clayton}, G.~C., \& {Mathis}, J.~S. 1989, \apj, 345, 245

\bibitem[{{Chabrier}(2003)}]{Chab2003}
{Chabrier}, G. 2003, \pasp, 115, 763

\bibitem[{Chabrier(2005)}]{Chab2005}
Chabrier, G. 2005, The Initial Mass Function 50 Years Later, ed Corelli E.,
  Palla F., \& Zinnecker H

\bibitem[{{Chen} {et~al.}(2013){Chen}, {Hickox}, {Alberts}, {Brodwin}, {Jones},
  {Murray}, {Alexander}, {Assef}, {Brown}, {Dey}, {Forman}, {Gorjian},
  {Goulding}, {Le Floc'h}, {Jannuzi}, {Mullaney}, \& {Pope}}]{Chen2013}
{Chen}, C.-T.~J., {Hickox}, R.~C., {Alberts}, S., {Brodwin}, M., {Jones}, C.,
  {Murray}, S.~S., {Alexander}, D.~M., {Assef}, R.~J., {Brown}, M.~J.~I.,
  {Dey}, A., {Forman}, W.~R., {Gorjian}, V., {Goulding}, A.~D., {Le Floc'h},
  E., {Jannuzi}, B.~T., {Mullaney}, J.~R., \& {Pope}, A. 2013, \apj, 773, 3

\bibitem[{{Croxall} {et~al.}(2013){Croxall}, {Smith}, {Brandl}, {Groves},
  {Kennicutt}, {Kreckel}, {Johnson}, {Pellegrini}, {Sandstrom}, {Walter},
  {Armus}, {Beir{\~a}o}, {Calzetti}, {Dale}, {Galametz}, {Hinz}, {Hunt},
  {Koda}, \& {Schinnerer}}]{Crox2013}
{Croxall}, K.~V., {Smith}, J.~D., {Brandl}, B.~R., {Groves}, B.~A.,
  {Kennicutt}, R.~C., {Kreckel}, K., {Johnson}, B.~D., {Pellegrini}, E.,
  {Sandstrom}, K.~M., {Walter}, F., {Armus}, L., {Beir{\~a}o}, P., {Calzetti},
  D., {Dale}, D.~A., {Galametz}, M., {Hinz}, J.~L., {Hunt}, L.~K., {Koda}, J.,
  \& {Schinnerer}, E. 2013, \apj, 777, 96

\bibitem[{{De Looze} {et~al.}(2014){De Looze}, {Cormier}, {Lebouteiller},
  {Madden}, {Baes}, {Bendo}, {Boquien}, {Boselli}, {Clements}, {Cortese},
  {Cooray}, {Galametz}, {Galliano}, {Gracia-Carpio}, {Isaak}, {Karczewski},
  {Parkin}, {Pellegrini}, {Remy-Ruyer}, {Spinoglio}, {Smith}, \&
  {Sturm}}]{DeLo2014}
{De Looze}, I., {Cormier}, D., {Lebouteiller}, V., {Madden}, S., {Baes}, M.,
  {Bendo}, G.~J., {Boquien}, M., {Boselli}, A., {Clements}, D.~L., {Cortese},
  L., {Cooray}, A., {Galametz}, M., {Galliano}, F., {Gracia-Carpio}, J.,
  {Isaak}, K., {Karczewski}, O.~L., {Parkin}, T.~J., {Pellegrini}, E.~W.,
  {Remy-Ruyer}, A., {Spinoglio}, L., {Smith}, M., \& {Sturm}, E. 2014, ArXiv
  e-prints

\bibitem[{{Elbaz} {et~al.}(2011){Elbaz}, {Dickinson}, {Hwang},
  {D{\'{\i}}az-Santos}, {Magdis}, {Magnelli}, {Le Borgne}, {Galliano},
  {Pannella}, {Chanial}, {Armus}, {Charmandaris}, {Daddi}, {Aussel}, {Popesso},
  {Kartaltepe}, {Altieri}, {Valtchanov}, {Coia}, {Dannerbauer}, {Dasyra},
  {Leiton}, {Mazzarella}, {Alexander}, {Buat}, {Burgarella}, {Chary}, {Gilli},
  {Ivison}, {Juneau}, {Le Floc'h}, {Lutz}, {Morrison}, {Mullaney}, {Murphy},
  {Pope}, {Scott}, {Brodwin}, {Calzetti}, {Cesarsky}, {Charlot}, {Dole},
  {Eisenhardt}, {Ferguson}, {F{\"o}rster Schreiber}, {Frayer}, {Giavalisco},
  {Huynh}, {Koekemoer}, {Papovich}, {Reddy}, {Surace}, {Teplitz}, {Yun}, \&
  {Wilson}}]{Elba2011}
{Elbaz}, D., {Dickinson}, M., {Hwang}, H.~S., {D{\'{\i}}az-Santos}, T.,
  {Magdis}, G., {Magnelli}, B., {Le Borgne}, D., {Galliano}, F., {Pannella},
  M., {Chanial}, P., {Armus}, L., {Charmandaris}, V., {Daddi}, E., {Aussel},
  H., {Popesso}, P., {Kartaltepe}, J., {Altieri}, B., {Valtchanov}, I., {Coia},
  D., {Dannerbauer}, H., {Dasyra}, K., {Leiton}, R., {Mazzarella}, J.,
  {Alexander}, D.~M., {Buat}, V., {Burgarella}, D., {Chary}, R.-R., {Gilli},
  R., {Ivison}, R.~J., {Juneau}, S., {Le Floc'h}, E., {Lutz}, D., {Morrison},
  G.~E., {Mullaney}, J.~R., {Murphy}, E., {Pope}, A., {Scott}, D., {Brodwin},
  M., {Calzetti}, D., {Cesarsky}, C., {Charlot}, S., {Dole}, H., {Eisenhardt},
  P., {Ferguson}, H.~C., {F{\"o}rster Schreiber}, N., {Frayer}, D.,
  {Giavalisco}, M., {Huynh}, M., {Koekemoer}, A.~M., {Papovich}, C., {Reddy},
  N., {Surace}, C., {Teplitz}, H., {Yun}, M.~S., \& {Wilson}, G. 2011, \aap,
  533, A119

\bibitem[{{Ellison} {et~al.}(2008){Ellison}, {Patton}, {Simard}, \&
  {McConnachie}}]{Elli2008}
{Ellison}, S.~L., {Patton}, D.~R., {Simard}, L., \& {McConnachie}, A.~W. 2008,
  \apjl, 672, L107

\bibitem[{{Gonzalez} {et~al.}(2013){Gonzalez}, {Sivanandam}, {Zabludoff}, \&
  {Zaritsky}}]{Gonz2013}
{Gonzalez}, A.~H., {Sivanandam}, S., {Zabludoff}, A.~I., \& {Zaritsky}, D.
  2013, \apj, 778, 14

\bibitem[{{Haxton} {et~al.}(2013){Haxton}, {Hamish Robertson}, \&
  {Serenelli}}]{Haxt2013}
{Haxton}, W.~C., {Hamish Robertson}, R.~G., \& {Serenelli}, A.~M. 2013, \araa,
  51, 21

\bibitem[{{Hoyle} {et~al.}(2012){Hoyle}, {Masters}, {Nichol}, {Jimenez}, \&
  {Bamford}}]{Hoyl2012}
{Hoyle}, B., {Masters}, K.~L., {Nichol}, R.~C., {Jimenez}, R., \& {Bamford},
  S.~P. 2012, \mnras, 423, 3478

\bibitem[{{Ilbert} {et~al.}(2013){Ilbert}, {McCracken}, {Le F{\`e}vre},
  {Capak}, {Dunlop}, {Karim}, {Renzini}, {Caputi}, {Boissier}, {Arnouts},
  {Aussel}, {Comparat}, {Guo}, {Hudelot}, {Kartaltepe}, {Kneib}, {Krogager},
  {Le Floc'h}, {Lilly}, {Mellier}, {Milvang-Jensen}, {Moutard}, {Onodera},
  {Richard}, {Salvato}, {Sanders}, {Scoville}, {Silverman}, {Taniguchi},
  {Tasca}, {Thomas}, {Toft}, {Tresse}, {Vergani}, {Wolk}, \& {Zirm}}]{Ilbe2013}
{Ilbert}, O., {McCracken}, H.~J., {Le F{\`e}vre}, O., {Capak}, P., {Dunlop},
  J., {Karim}, A., {Renzini}, M.~A., {Caputi}, K., {Boissier}, S., {Arnouts},
  S., {Aussel}, H., {Comparat}, J., {Guo}, Q., {Hudelot}, P., {Kartaltepe}, J.,
  {Kneib}, J.~P., {Krogager}, J.~K., {Le Floc'h}, E., {Lilly}, S., {Mellier},
  Y., {Milvang-Jensen}, B., {Moutard}, T., {Onodera}, M., {Richard}, J.,
  {Salvato}, M., {Sanders}, D.~B., {Scoville}, N., {Silverman}, J.~D.,
  {Taniguchi}, Y., {Tasca}, L., {Thomas}, R., {Toft}, S., {Tresse}, L.,
  {Vergani}, D., {Wolk}, M., \& {Zirm}, A. 2013, \aap, 556, A55

\bibitem[{{Kauffmann} {et~al.}(2003{\natexlab{a}}){Kauffmann}, {Heckman},
  {Tremonti}, {Brinchmann}, {Charlot}, {White}, {Ridgway}, {Brinkmann},
  {Fukugita}, {Hall}, {Ivezi{\'c}}, {Richards}, \& {Schneider}}]{Kauf2003a}
{Kauffmann}, G., {Heckman}, T.~M., {Tremonti}, C., {Brinchmann}, J., {Charlot},
  S., {White}, S.~D.~M., {Ridgway}, S.~E., {Brinkmann}, J., {Fukugita}, M.,
  {Hall}, P.~B., {Ivezi{\'c}}, {\v Z}., {Richards}, G.~T., \& {Schneider},
  D.~P. 2003{\natexlab{a}}, \mnras, 346, 1055

\bibitem[{{Kauffmann} {et~al.}(2003{\natexlab{b}}){Kauffmann}, {Heckman},
  {White}, {Charlot}, {Tremonti}, {Brinchmann}, {Bruzual}, {Peng}, {Seibert},
  {Bernardi}, {Blanton}, {Brinkmann}, {Castander}, {Cs{\'a}bai}, {Fukugita},
  {Ivezic}, {Munn}, {Nichol}, {Padmanabhan}, {Thakar}, {Weinberg}, \&
  {York}}]{Kauf2003b}
{Kauffmann}, G., {Heckman}, T.~M., {White}, S.~D.~M., {Charlot}, S.,
  {Tremonti}, C., {Brinchmann}, J., {Bruzual}, G., {Peng}, E.~W., {Seibert},
  M., {Bernardi}, M., {Blanton}, M., {Brinkmann}, J., {Castander}, F.,
  {Cs{\'a}bai}, I., {Fukugita}, M., {Ivezic}, Z., {Munn}, J.~A., {Nichol},
  R.~C., {Padmanabhan}, N., {Thakar}, A.~R., {Weinberg}, D.~H., \& {York}, D.
  2003{\natexlab{b}}, \mnras, 341, 33

\bibitem[{{Kennicutt}(1998{\natexlab{a}})}]{Kenn1998a}
{Kennicutt}, Jr., R.~C. 1998{\natexlab{a}}, \araa, 36, 189

\bibitem[{{Kennicutt}(1998{\natexlab{b}})}]{Kenn1998b}
---. 1998{\natexlab{b}}, \apj, 498, 541

\bibitem[{{Kroupa}(2001)}]{Krou2001}
{Kroupa}, P. 2001, \mnras, 322, 231

\bibitem[{{Krumholz} {et~al.}(2012){Krumholz}, {Dekel}, \& {McKee}}]{Krum2012}
{Krumholz}, M.~R., {Dekel}, A., \& {McKee}, C.~F. 2012, \apj, 745, 69

\bibitem[{{Krumholz} {et~al.}(2009){Krumholz}, {McKee}, \&
  {Tumlinson}}]{Krum2009}
{Krumholz}, M.~R., {McKee}, C.~F., \& {Tumlinson}, J. 2009, \apj, 699, 850

\bibitem[{{Lara-L{\'o}pez} {et~al.}(2010){Lara-L{\'o}pez}, {Cepa},
  {Bongiovanni}, {P{\'e}rez Garc{\'{\i}}a}, {Ederoclite}, {Casta{\~n}eda},
  {Fern{\'a}ndez Lorenzo}, {Povi{\'c}}, \& {S{\'a}nchez-Portal}}]{Lara2010}
{Lara-L{\'o}pez}, M.~A., {Cepa}, J., {Bongiovanni}, A., {P{\'e}rez
  Garc{\'{\i}}a}, A.~M., {Ederoclite}, A., {Casta{\~n}eda}, H., {Fern{\'a}ndez
  Lorenzo}, M., {Povi{\'c}}, M., \& {S{\'a}nchez-Portal}, M. 2010, \aap, 521,
  L53

\bibitem[{{Lovisari} {et~al.}(2011){Lovisari}, {Schindler}, \&
  {Kapferer}}]{Lovi2011}
{Lovisari}, L., {Schindler}, S., \& {Kapferer}, W. 2011, \aap, 528, A60

\bibitem[{{Mannucci} {et~al.}(2010){Mannucci}, {Cresci}, {Maiolino}, {Marconi},
  \& {Gnerucci}}]{Mann2010}
{Mannucci}, F., {Cresci}, G., {Maiolino}, R., {Marconi}, A., \& {Gnerucci}, A.
  2010, \mnras, 408, 2115

\bibitem[{{Meynet} \& {Maeder}(2002)}]{Meyn2002}
{Meynet}, G., \& {Maeder}, A. 2002, \aap, 390, 561

\bibitem[{{Oppenheimer} \& {Dav{\'e}}(2008)}]{Oppe2008}
{Oppenheimer}, B.~D., \& {Dav{\'e}}, R. 2008, \mnras, 387, 577

\bibitem[{{Oppenheimer} {et~al.}(2010){Oppenheimer}, {Dav{\'e}}, {Kere{\v s}},
  {Fardal}, {Katz}, {Kollmeier}, \& {Weinberg}}]{Oppe2010}
{Oppenheimer}, B.~D., {Dav{\'e}}, R., {Kere{\v s}}, D., {Fardal}, M., {Katz},
  N., {Kollmeier}, J.~A., \& {Weinberg}, D.~H. 2010, \mnras, 406, 2325

\bibitem[{{Parravano} {et~al.}(2011){Parravano}, {McKee}, \&
  {Hollenbach}}]{Parr2011}
{Parravano}, A., {McKee}, C.~F., \& {Hollenbach}, D.~J. 2011, \apj, 726, 27

\bibitem[{{Petropoulou} {et~al.}(2012){Petropoulou}, {V{\'{\i}}lchez}, \&
  {Iglesias-P{\'a}ramo}}]{Petr2012}
{Petropoulou}, V., {V{\'{\i}}lchez}, J., \& {Iglesias-P{\'a}ramo}, J. 2012,
  \apj, 749, 133

\bibitem[{{Rosario} {et~al.}(2013){Rosario}, {Santini}, {Lutz}, {Netzer},
  {Bauer}, {Berta}, {Magnelli}, {Popesso}, {Alexander}, {Brandt}, {Genzel},
  {Maiolino}, {Mullaney}, {Nordon}, {Saintonge}, {Tacconi}, \&
  {Wuyts}}]{Rosa2013}
{Rosario}, D.~J., {Santini}, P., {Lutz}, D., {Netzer}, H., {Bauer}, F.~E.,
  {Berta}, S., {Magnelli}, B., {Popesso}, P., {Alexander}, D.~M., {Brandt},
  W.~N., {Genzel}, R., {Maiolino}, R., {Mullaney}, J.~R., {Nordon}, R.,
  {Saintonge}, A., {Tacconi}, L., \& {Wuyts}, S. 2013, \apj, 771, 63

\bibitem[{{Salpeter}(1959)}]{Salp1959}
{Salpeter}, E.~E. 1959, \apj, 129, 608

\bibitem[{{S{\'a}nchez} {et~al.}(2013){S{\'a}nchez}, {Rosales-Ortega},
  {Jungwiert}, {Iglesias-P{\'a}ramo}, {V{\'{\i}}lchez}, {Marino}, {Walcher},
  {Husemann}, {Mast}, {Monreal-Ibero}, {Cid Fernandes}, {P{\'e}rez},
  {Gonz{\'a}lez Delgado}, {Garc{\'{\i}}a-Benito}, {Galbany}, {van de Ven},
  {Jahnke}, {Flores}, {Bland-Hawthorn}, {L{\'o}pez-S{\'a}nchez}, {Stanishev},
  {Miralles-Caballero}, {D{\'{\i}}az}, {S{\'a}nchez-Blazquez}, {Moll{\'a}},
  {Gallazzi}, {Papaderos}, {Gomes}, {Gruel}, {P{\'e}rez}, {Ruiz-Lara},
  {Florido}, {de Lorenzo-C{\'a}ceres}, {Mendez-Abreu}, {Kehrig}, {Roth},
  {Ziegler}, {Alves}, {Wisotzki}, {Kupko}, {Quirrenbach}, {Bomans}, \& {Califa
  Collaboration}}]{Sanc2013}
{S{\'a}nchez}, S.~F., {Rosales-Ortega}, F.~F., {Jungwiert}, B.,
  {Iglesias-P{\'a}ramo}, J., {V{\'{\i}}lchez}, J.~M., {Marino}, R.~A.,
  {Walcher}, C.~J., {Husemann}, B., {Mast}, D., {Monreal-Ibero}, A., {Cid
  Fernandes}, R., {P{\'e}rez}, E., {Gonz{\'a}lez Delgado}, R.,
  {Garc{\'{\i}}a-Benito}, R., {Galbany}, L., {van de Ven}, G., {Jahnke}, K.,
  {Flores}, H., {Bland-Hawthorn}, J., {L{\'o}pez-S{\'a}nchez}, A.~R.,
  {Stanishev}, V., {Miralles-Caballero}, D., {D{\'{\i}}az}, A.~I.,
  {S{\'a}nchez-Blazquez}, P., {Moll{\'a}}, M., {Gallazzi}, A., {Papaderos}, P.,
  {Gomes}, J.~M., {Gruel}, N., {P{\'e}rez}, I., {Ruiz-Lara}, T., {Florido}, E.,
  {de Lorenzo-C{\'a}ceres}, A., {Mendez-Abreu}, J., {Kehrig}, C., {Roth},
  M.~M., {Ziegler}, B., {Alves}, J., {Wisotzki}, L., {Kupko}, D.,
  {Quirrenbach}, A., {Bomans}, D., \& {Califa Collaboration}. 2013, \aap, 554,
  A58

\bibitem[{{Savaglio} {et~al.}(2005){Savaglio}, {Glazebrook}, {Le Borgne},
  {Juneau}, {Abraham}, {Chen}, {Crampton}, {McCarthy}, {Carlberg}, {Marzke},
  {Roth}, {J{\o}rgensen}, \& {Murowinski}}]{Sava2005}
{Savaglio}, S., {Glazebrook}, K., {Le Borgne}, D., {Juneau}, S., {Abraham},
  R.~G., {Chen}, H.-W., {Crampton}, D., {McCarthy}, P.~J., {Carlberg}, R.~G.,
  {Marzke}, R.~O., {Roth}, K., {J{\o}rgensen}, I., \& {Murowinski}, R. 2005,
  \apj, 635, 260

\bibitem[{{Schaller} {et~al.}(1992){Schaller}, {Schaerer}, {Meynet}, \&
  {Maeder}}]{Scha1992}
{Schaller}, G., {Schaerer}, D., {Meynet}, G., \& {Maeder}, A. 1992, \aaps, 96,
  269

\bibitem[{{Schmidt}(1959)}]{Schm1959}
{Schmidt}, M. 1959, \apj, 129, 243

\bibitem[{{Singh} {et~al.}(2013){Singh}, {van de Ven}, {Jahnke}, {Lyubenova},
  {Falc{\'o}n-Barroso}, {Alves}, {Cid Fernandes}, {Galbany},
  {Garc{\'{\i}}a-Benito}, {Husemann}, {Kennicutt}, {Marino}, {M{\'a}rquez},
  {Masegosa}, {Mast}, {Pasquali}, {S{\'a}nchez}, {Walcher}, {Wild}, {Wisotzki},
  \& {Ziegler}}]{Sing2013}
{Singh}, R., {van de Ven}, G., {Jahnke}, K., {Lyubenova}, M.,
  {Falc{\'o}n-Barroso}, J., {Alves}, J., {Cid Fernandes}, R., {Galbany}, L.,
  {Garc{\'{\i}}a-Benito}, R., {Husemann}, B., {Kennicutt}, R.~C., {Marino},
  R.~A., {M{\'a}rquez}, I., {Masegosa}, J., {Mast}, D., {Pasquali}, A.,
  {S{\'a}nchez}, S.~F., {Walcher}, J., {Wild}, V., {Wisotzki}, L., \&
  {Ziegler}, B. 2013, \aap, 558, A43

\bibitem[{{Torres-Papaqui} {et~al.}(2012){Torres-Papaqui}, {Coziol},
  {Ortega-Minakata}, \& {Neri-Larios}}]{Torr2012}
{Torres-Papaqui}, J.~P., {Coziol}, R., {Ortega-Minakata}, R.~A., \&
  {Neri-Larios}, D.~M. 2012, \apj, 754, 144

\bibitem[{{Tremonti} {et~al.}(2004){Tremonti}, {Heckman}, {Kauffmann},
  {Brinchmann}, {Charlot}, {White}, {Seibert}, {Peng}, {Schlegel}, {Uomoto},
  {Fukugita}, \& {Brinkmann}}]{Trem2004}
{Tremonti}, C.~A., {Heckman}, T.~M., {Kauffmann}, G., {Brinchmann}, J.,
  {Charlot}, S., {White}, S.~D.~M., {Seibert}, M., {Peng}, E.~W., {Schlegel},
  D.~J., {Uomoto}, A., {Fukugita}, M., \& {Brinkmann}, J. 2004, \apj, 613, 898

\bibitem[{{Wong} {et~al.}(2012){Wong}, {Schawinski}, {Kaviraj}, {Masters},
  {Nichol}, {Lintott}, {Keel}, {Darg}, {Bamford}, {Andreescu}, {Murray},
  {Raddick}, {Szalay}, {Thomas}, \& {Vandenberg}}]{Wong2012}
{Wong}, O.~I., {Schawinski}, K., {Kaviraj}, S., {Masters}, K.~L., {Nichol},
  R.~C., {Lintott}, C., {Keel}, W.~C., {Darg}, D., {Bamford}, S.~P.,
  {Andreescu}, D., {Murray}, P., {Raddick}, M.~J., {Szalay}, A., {Thomas}, D.,
  \& {Vandenberg}, J. 2012, \mnras, 420, 1684

\bibitem[{{Wright}(2006)}]{Wrig2006}
{Wright}, E.~L. 2006, \pasp, 118, 1711

\bibitem[{{Wu} \& {Zhang}(2013)}]{WuYZ2013}
{Wu}, Y.-Z., \& {Zhang}, S.-N. 2013, \mnras, 436, 934

\bibitem[{{Yates} {et~al.}(2012){Yates}, {Kauffmann}, \& {Guo}}]{Yate2012}
{Yates}, R.~M., {Kauffmann}, G., \& {Guo}, Q. 2012, \mnras, 422, 215

\end{thebibliography}
\end{document}